\documentclass[11pt]{article}
\DeclareMathAlphabet{\scr}{U}{rsfs}{m}{n}
\usepackage{latexsym}
\usepackage{epsfig}
\usepackage[mathscr]{eucal}
\usepackage{amsfonts}
\usepackage{amscd}
\usepackage{amsmath}
\usepackage{array}
\usepackage{amssymb}
\usepackage{colordvi}
\usepackage{enumerate}
\usepackage{graphicx}
\usepackage{booktabs}
\usepackage[footnotesize]{caption}
\usepackage{fancyhdr} 
\usepackage{pdfpages}
\usepackage{slashed}
\usepackage{tabularx}
\usepackage{longtable}
\usepackage{array}
\usepackage{hyperref}
\usepackage{relsize}
\usepackage{color}
\usepackage[merge,numbers,compress]{natbib}
\usepackage{rotating}
\usepackage{cleveref}
\usepackage{subfig}
\crefformat{footnote}{#2\footnotemark[#1]#3}

\hypersetup{
	colorlinks=true,        % false: boxed links; true: colored links
	linkcolor=blue,          % color of internal links
	citecolor=black,        % color of links to bibliography
	filecolor=red,      % color of file links
	urlcolor=cyan           % color of external links
}

\definecolor{ourbrown}{RGB}{155,100,15}
\definecolor{ourcyan}{RGB}{20,165,165}
\definecolor{ourpurple}{RGB}{145,0,140}
\definecolor{darkorange}{RGB}{225,100,0}
\definecolor{darkgreen}{RGB}{0,170,0}
\definecolor{darkgray}{RGB}{80,80,80}

\newcommand{\newc}{\newcommand}
\newc{\EW}{electroweak\;}
\newc{\DM}{dark matter\;}
\newc{\SM}{standard model\;}
\newc{\GEV}{\text{GeV}}
\newc{\KK}{Kaluza-Klein\;}
\newc{\ff}{fragmentation function\;}
\newc{\be}{\begin{equation}}
\newc{\ee}{\end{equation}}
\newc{\bi}{\begin{itemize}}
\newc{\ei}{\end{itemize}}
\newc{\benu}{\begin{enumerate}}
\newc{\eenu}{\end{enumerate}}
\newc{\bc}{\begin{center}}
\newc{\ec}{\end{center}}
\newc{\bfig}{\begin{figure}}
\newc{\efig}{\end{figure}}
\newc{\neutone}{\tilde{\chi}^0_1}
\newc{\sigmav}{\langle\sigma v \rangle}
\newc{\lamhs}{\lambda_{H\!S}}
\newc{\logJ}{\log (J/J_{\text{nom}})}
\newc{\sigJ}{\sigma_{\log\!J}}
\DeclareMathSymbol{\Gamma}{\mathalpha}{letters}{"00}
\DeclareMathSymbol{\Delta}{\mathalpha}{letters}{"01}
\DeclareMathSymbol{\Theta}{\mathalpha}{letters}{"02}
\DeclareMathSymbol{\Lambda}{\mathalpha}{letters}{"03}
\DeclareMathSymbol{\Xi}{\mathalpha}{letters}{"04}
\DeclareMathSymbol{\Pi}{\mathalpha}{letters}{"05}
\DeclareMathSymbol{\Sigma}{\mathalpha}{letters}{"06}
\DeclareMathSymbol{\Upsilon}{\mathalpha}{letters}{"07}
\DeclareMathSymbol{\Phi}{\mathalpha}{letters}{"08}
\DeclareMathSymbol{\Psi}{\mathalpha}{letters}{"09}
\DeclareMathSymbol{\Omega}{\mathalpha}{letters}{"0A}
\DeclareMathSymbol{\varGamma}{\mathalpha}{operators}{"00}
\DeclareMathSymbol{\varDelta}{\mathalpha}{operators}{"01}
\DeclareMathSymbol{\varTheta}{\mathalpha}{operators}{"02}
\DeclareMathSymbol{\varLambda}{\mathalpha}{operators}{"03}
\DeclareMathSymbol{\varXi}{\mathalpha}{operators}{"04}
\DeclareMathSymbol{\varPi}{\mathalpha}{operators}{"05}
\DeclareMathSymbol{\varSigma}{\mathalpha}{operators}{"06}
\DeclareMathSymbol{\varUpsilon}{\mathalpha}{operators}{"07}
\DeclareMathSymbol{\varPhi}{\mathalpha}{operators}{"08}
\DeclareMathSymbol{\varPsi}{\mathalpha}{operators}{"09}
\DeclareMathSymbol{\varOmega}{\mathalpha}{operators}{"0A}

\renewcommand{\vec}[1]{\boldsymbol{#1}}
\newcommand{\D}{\mathrm{d}}

\newcommand{\hn}{h^0}
\newcommand{\Hn}{H^0}
\newcommand{\An}{A^0}
\newcommand{\Hp}{H^\pm}

\newcommand{\lam}{\lambda}
\newcommand{\lamL}{\lambda_L}
\newcommand{\HO}{H^0}
\newcommand{\sv}{\langle\sigma v \rangle}

\begin{document}

\title{\hfill ~\\[-30mm]
\phantom{h} \hfill\mbox{\small TTK-17-12} 
\\[1cm]
\vspace{13mm}  
\textbf{
The inert doublet model in the light of Fermi-LAT gamma-ray data:
a global fit analysis
}
\\ \vspace{1ex}
}
\date{}
\author{
Benedikt Eiteneuer$^a$\footnote{E-mail: \texttt{eiteneuer@physik.rwth-aachen.de}}\;,
Andreas Goudelis$^b$\footnote{E-mail: \texttt{andreas.goudelis@lpthe.jussieu.fr}}\;, and
Jan Heisig$^a$\footnote{E-mail: \texttt{heisig@physik.rwth-aachen.de}}\\[9mm]
{\it $^a$Institute for Theoretical Particle Physics and Cosmology,}\\ {\it RWTH Aachen University, 52056 Aachen, Germany}\\[2mm]
{\it $^b$Laboratoire de Physique Th\'eorique et Hautes Energies (LPTHE),}\\
{\it UMR 7589 CNRS \& UPMC, 4 Place Jussieu, F-75252, Paris, France}
}

\maketitle

\begin{abstract}
We perform a global fit within the inert doublet model taking into account experimental observables from colliders, direct and indirect dark matter searches and theoretical constraints. In particular, we consider recent results from searches for dark matter annihilation-induced gamma-rays in dwarf spheroidal galaxies and relax the assumption that the inert doublet model should account for the entire dark matter in the Universe. We moreover study in how far the model is compatible with a possible dark matter explanation of the so-called Galactic center excess. We find two distinct parameter space regions that are consistent with existing constraints and can simultaneously explain the excess: One with dark matter masses near the Higgs resonance and one around 72 GeV where dark matter annihilates predominantly into pairs of virtual electroweak gauge bosons via the four-vertex arising from the inert doublet's kinetic term. We briefly discuss future prospects to probe these scenarios.
\end{abstract}
\thispagestyle{empty}
\vfill
\newpage
\setcounter{page}{1}

\tableofcontents

%===================================================================
\section{Introduction}\label{sec:intro}
%===================================================================

The nature of dark matter (DM), the existence of which is corroborated by observations over a wide range of physical scales in the Universe, is one of the 
most important open questions in contemporary fundamental physics. An explanation in terms of a weakly interacting 
massive particle (WIMP) is an attractive possibility which has motivated an enormous experimental effort.
Indirect detection experiments have reached sensitivity to the thermal self-annihilation cross section for DM 
masses around the electroweak scale and direct detection experiments have substantially improved 
the limits on WIMP-nucleon scattering over the past few years. Interpreting these results in terms
of well-motivated theoretical models is hence an important task in order to pinpoint the nature of DM\@.

The inert doublet model (IDM)\footnote{%
For recent accounts see, \emph{e.g.}~\cite{Ilnicka:2015jba,Belyaev:2016lok}.} 
is among the simplest new physics models, supplementing
the standard model with an additional complex scalar field that transforms as a doublet under $SU(2)_L$ 
and is odd under a discrete ${\cal{Z}}_2$ symmetry, all standard model fields being taken to 
be even. Despite its simplicity the IDM has a rich and versatile phenomenology: it can affect electroweak 
symmetry breaking~\cite{Deshpande:1977rw,Barbieri:2006dq,Hambye:2007vf,Khan:2015ipa,
Swiezewska:2015paa,Plascencia:2015xwa}, give rise to interesting, observable effects at colliders~\cite{Belyaev:2016lok,Lundstrom:2008ai,
Pierce:2007ut,Cao:2007rm,Dolle:2009ft,Miao:2010rg,Arhrib:2012ia,Gustafsson:2012aj,Swiezewska:2012eh,
Goudelis:2013uca,Belanger:2015kga}, modifiy electroweak baryogenesis~\cite{Chowdhury:2011ga,Borah:2012pu}, 
play a role in the generation of neutrino masses \cite{Ma:2006km} and, as being the focus of this work, 
it contains a WIMP that can account for the observed DM
in the Universe with observable signatures in direct and indirect detection experiments~\cite{Barbieri:2006dq,Goudelis:2013uca,
LopezHonorez:2006gr,Hambye:2009pw,Honorez:2010re,LopezHonorez:2010tb,Gustafsson:2007pc,
Agrawal:2008xz,Andreas:2009hj,Nezri:2009jd,Garcia-Cely:2013zga,Garcia-Cely:2015khw,Queiroz:2015utg,Benito:2016kyp}.
The versatility of the IDM as a DM model introduces a fair amount of freedom to accommodate measurements and constraints from various observables,
making it a non-trivial task to unfold the data and extract information about the physical parameters of the model. 

In this regard, global fit techniques are of central importance. They enable the systematic study of the impact of a large number of experimental measurements while fully accounting for systematic uncertainties 
that affect astrophysical observables such as the DM density profile in the inner galaxy.
In this paper we perform a detailed numerical fit within the IDM using  \textsc{MultiNest}~\cite{Feroz:2008xx,Feroz:2013hea},
which allows us to 
comprehensively explore the model's parameter space. Furthermore,
instead of demanding that the IDM dark matter candidate should account for the entire DM abundance in the Universe,
we follow a more general approach allowing for an unspecified additional DM component 
to contribute subdominantly or even dominantly to the total DM density. Introducing the fractional density of IDM dark matter as a free 
parameter in the fit enables us to extract information about the amount of DM that can be accommodated within the model. 

We consider two setups. On the one hand,
we fit a set of well-established observables: The DM relic density measured by Planck~\cite{Ade:2015xua}, direct detection constraints set by LUX~\cite{Akerib:2016vxi}, 
indirect detection constraints from the observation of dwarf spheroidal galaxies 
set by Fermi-LAT~\cite{Ackermann:2015zua,Fermi-LAT:2016uux} as well as the Higgs mass measured at the LHC~\cite{Aad:2015zhl}, constraints from
invisible Higgs decays~\cite{Aad:2015pla}, constraints
from electroweak precision tests~\cite{Baak:2014ora} and theoretical bounds from unitarity, 
perturbativity and vacuum stability.

On the other hand, over the past few years the Fermi-LAT gamma-ray data revealed an unexpected
Galactic bulge emission component -- the ``Galactic center excess'' (GCE)~\cite{Goodenough:2009gk,Hooper:2010mq,Hooper:2011ti,Abazajian:2012pn,Hooper:2013rwa,Gordon:2013vta,Abazajian:2014fta,Daylan:2014rsa,Zhou:2014lva,Calore:2014xka,Agrawal:2014oha,Calore:2014nla,Huang:2015rlu,TheFermi-LAT:2015kwa,Karwin:2016tsw}. 
Although various  astrophysical explanations have been 
proposed~\cite{Petrovic:2014uda,Petrovic:2014xra,Yuan:2014rca,Cholis:2015dea} (for statistical
approaches to test the origin of the signal see further discussions in~\cite{Lee:2014mza,Lee:2015fea,Bartels:2015aea,McDermott:2015ydv,Horiuchi:2016zwu}), it is intriguing that 
the strength as well as spectral and morphological 
properties of the excess are compatible with a signal from DM annihilation with thermal cross section 
and a DM mass $m_\text{DM}\lesssim100\,\text{GeV}$.
Given the complexity of the Galactic center as an astrophysical environment, dwarf spheroidal galaxies (dSphs) are a 
much cleaner target for DM searches in gamma-rays 
as their dynamical and chemical properties suggest larger mass-to-light ratios.
Hence they provide an important test of the DM interpretation of the GCE\@.
Searches for a gamma-ray excess associated with dSphs have been
performed and the sensitivity is competitive 
with that of other targets such as the Galactic center. 
Interestingly, slight excesses (each with a $\sim2\sigma$ local significance) have been found in
four of the recently discovered dSph targets~\cite{Geringer-Sameth:2015lua,Li:2015kag,Fermi-LAT:2016uux}
which are roughly compatible with a DM explanation of the GCE\@. Given these hints, as a second step in this paper we include the
GCE in addition to the observables mentioned above in our global fit.\footnote{For other attempts to perform global fits to the GCE within UV-complete models see, \textit{e.g.}~\cite{Butter:2015fqa,Butter:2016tjc}.}
Note also that similarly, an excess which appears to be compatible with 
a signal from DM annihilations as well as with the GCE
itself has recently been reported \cite{Cuoco:2016eej} in the AMS-02 antiproton data~\cite{Aguilar:2016kjl} 
(see~\cite{Cui:2016ppb} for a similar result using the boron over carbon ratio~\cite{Aguilar:2016vqr}).
An interpretation in terms of individual DM annihilation channels but also in terms of the singlet scalar
DM model was presented in \cite{Cuoco:2017rxb}.
Although an analysis of the antiproton flux measurements falls beyond the scope of the present paper, these hints provide additional motivation to
study whether the IDM can accommodate the GCE.

The paper is organized as follows: in section~\ref{sec:model} we briefly introduce the IDM\@.
The various observables included in our fitting procedure, along with the method followed in
order to sample the IDM parameter space, are detailed in section~\ref{sec:constraints}. Section~\ref{sec:results} contains our 
main results and a discussion on the future sensitivities of upcoming experiments to the best-fit parameter regions. We conclude in 
section~\ref{sec:summary}.

%===================================================================
\section{The inert doublet model}\label{sec:model}
%===================================================================

The IDM is a special case of a two-Higgs doublet model
in which an exact ${\cal{Z}}_2$ discrete symmetry is imposed on the Lagrangian, 
under which all standard model fields (including the usual Higgs doublet $H$) 
are taken to be even, whereas the second scalar doublet $\Phi$ is taken to be odd. 
With respect to the standard model Lagrangian, the only modifications consist of the introduction of
gauge kinetic terms for $\Phi$ and an additional piece in the scalar potential, which in the IDM reads
\begin{equation}
	V ~=~ \mu_1^2 |H|^2  + \mu_2^2|\Phi|^2 + \lambda_1 |H|^4+ \lambda_2 |\Phi|^4 + \lambda_3 |H|^2| \Phi|^2
		+ \lambda_4 |H^\dagger\Phi|^2 + \frac{\lambda_5}{2} \Bigl[ (H^\dagger\Phi)^2 + \mathrm{h.c.} \Bigr].
\label{Eq:TreePotential}
\end{equation}
Upon electroweak symmetry breaking, the two scalar doublets can be expanded in component fields as
\begin{equation}
	H ~=~ \left( \begin{array}{c} G^+ \\ \frac{1}{\sqrt{2}}\left(v+\hn+\mathrm{i}G^0\right) \end{array} \right),
	\qquad
	\Phi ~=~ \left( \begin{array}{c} H^+\\ \frac{1}{\sqrt{2}}\left(H^0+\mathrm{i}A^0\right) \end{array} \right),
\end{equation}
where $v \simeq 246$ GeV is the usual Higgs field vacuum expectation value and $G$ are the Goldstone bosons. The model contains five physical scalar states with masses given by
\begin{align}\label{eq:physmasses}
	m_{\hn}^2 &= \mu_1^2 + 3 \lambda_1 v^2, \\ \nonumber
	m_{\Hn}^2 &= \mu_2^2 + \lambda_L v^2, \\ \nonumber
	m_{\An}^2 &= \mu_2^2 + \lambda_S v^2, \\ \nonumber
	m_{\Hp}^2 &= \mu_2^2 + \frac{1}{2} \lambda_3 v^2,
\end{align}
where, following common conventions, we have defined
\begin{align}\label{eq:lambdaLS}
	\lambda_{L,S} & = \frac{1}{2} \left( \lambda_3 + \lambda_4 \pm \lambda_5 \right).
\end{align}
These parameters correspond to the coupling of a pair of $\Hn$, $\An$ states, respectively, to the Higgs boson. All in all, the IDM is characterized by six free parameters:
\begin{equation}
\left\{ \lam_1, ~~ \lam_2, ~~ \lam_3, ~~ \lam_4, ~~ \lam_5, ~~ \mu_2 \right\},
\label{eq:lambdas}
\end{equation}
which, using equations \eqref{eq:physmasses}, \eqref{eq:lambdaLS} and the potential minimization condition $\left.\left(\partial V/ \partial \hn\right)\right|_{\hn = 0} = 0$ can be traded for the physically more intuitive set of parameters
\begin{equation}
 	\left\{ m_{\hn}, ~~ m_{\Hn}, ~~ m_{\An}, ~~ m_{\Hp}, ~~ \lambda_L, ~~ \lambda_2 \right\},
	\label{eq:masses}
\end{equation} 
with $m_{\hn} \simeq 125.09$ GeV \cite{pdg} being the measured Higgs boson mass. 

The discrete symmetry imposed on the Lagrangian renders the lightest component of the inert doublet stable. If, moreover, this lightest ${\cal{Z}}_2$-odd 
particle is neutral ($m_{\Hn/\An} < m_{\Hp}$), it can play
the role of a DM candidate. In fact, the IDM is perhaps the simplest model in which the observed DM abundance in the Universe can be obtained
through all ways that typically characterize WIMP models: adjusting couplings, approaching or taking distance from resonances/thresholds and coannihilation. Note also that
the DM phenomenology of $\Hn$ and $\An$ is identical. In this respect, for simplicity in what follows we will consistently adopt the choice $m_{\Hn} < m_{\An}$. 
A more detailed description of the IDM phenomenology will be presented in the following sections.

%===================================================================
\section{Constraints and global fit settings}\label{sec:constraints}
%===================================================================

Various aspects of the IDM phenomenology have been studied in the literature. The model was first proposed as a DM model in \cite{Barbieri:2006dq}, while 
its predicted relic abundance was analysed in more detail in \cite{LopezHonorez:2006gr,Hambye:2009pw,Honorez:2010re,LopezHonorez:2010tb}. Direct detection constraints
were first considered in \cite{Barbieri:2006dq} (as well as in most subsequent studies), whereas indirect detection has been studied for continuum gamma-rays \cite{LopezHonorez:2006gr,Queiroz:2015utg},
spectral features \cite{Gustafsson:2007pc,Garcia-Cely:2013zga,Garcia-Cely:2015khw}, antimatter \cite{Nezri:2009jd} and neutrinos \cite{Agrawal:2008xz,Andreas:2009hj}. Other than the DM abundance, 
direct detection is known to impose extremely strong constraints on the IDM parameter space whereas currently the strongest indirect detection bounds stem from gamma-ray searches in 
dwarf spheroidal galaxies. Besides, the region of parameter space where important gamma-ray lines could be expected is severely bound by other observations. 

In \cite{Goudelis:2013uca} it was shown that the invisible Higgs width imposes strong bounds on the $\lambda_L$ coupling if $m_{\Hn} < m_{\hn}/2$, whereas the masses
of the next-to-lightest states can be constrained from LEP-II searches for neutralinos \cite{Lundstrom:2008ai} and charginos \cite{Pierce:2007ut}. The new states can also 
induce contributions to the $S$ and $T$ electroweak parameters, as first pointed out in \cite{Barbieri:2006dq}. Constraints from searches for dileptons along with missing
transverse energy at the LHC were first proposed in \cite{Dolle:2009ft} and have been analysed in \cite{Belanger:2015kga}. Although during Run-II they will offer the
opportunity to test an interesting part of the IDM parameter space, at present their impact is subleading with respect to other searches. Lastly, a minimal 
set of requirements must be imposed on the parameter choices in order to ensure that the electroweak vacuum is stable (for detailed analyses \textit{cf.}~\cite{Khan:2015ipa,Swiezewska:2015paa,Goudelis:2013uca}) and that perturbative calculations make sense, also in the sense of perturbative unitarity of the scattering 
matrix.

We now proceed to discuss these constraints and the way they are incorporated in our global likelihood fit in more detail.

%------------------------------------------------------------------------
\subsection{Dark matter observables}
%------------------------------------------------------------------------

%- - - - - - - - - - - - - - - - - - - - - - - - - - - - - - - - - - - - - -
\subsubsection{Relic density}  \label{sec:reldens}
%- - - - - - - - - - - - - - - - - - - - - - - - - - - - - - - - - - - - - -

Assuming a standard cosmological history allows us to link the relic $\Hn$ density from thermal freeze-out, $\Omega h^2|_\text{IDM}$, to the DM density measured  by Planck, 
$\Omega h^2|_\text{Planck} = 0.1198\pm 0.0015$~\cite{Ade:2015xua}. 
In this study we allow for the possibility that the dark sector might be comprised of more than 
one DM component by introducing the fraction of the DM density predicted from the IDM over the total DM density in the Universe
\begin{equation}
 R\equiv\frac{\rho_\text{IDM}}{\rho_{\text{total}}}\,
\end{equation}
as a free (astrophysical) parameter.
We assume that the clustering properties and, hence, 
the density profiles of the IDM and non-IDM DM components behave 
sufficiently similarly so that they constitute the same fraction $R$ of DM on the different scales which are relevant for the various DM observables considered here. 
Then, the total DM density is given by
\begin{equation}
\Omega h^2|_\text{DM,\,total}=\frac{\Omega h^2|_\text{IDM}}{R}\,.
\end{equation}
The $\Hn$ relic density $\Omega h^2|_\text{IDM}$ has been calculated using \textsc{micrOMEGAs}~\cite{Belanger:2014vza}, thanks to an implementation of the model
in the \textsc{FeynRules} package \cite{Alloul:2013bka}. 
Our computations take into account 3-body final state contributions to the total DM annihilation cross section, which can be extremely important in some regions of parameter space.
\\
\\
We compute the $\chi^2$ contribution for the relic density via
\begin{equation}
\chi_\Omega^2 = \frac{\left(\Omega h^2|_\text{DM,\,total}-\Omega h^2|_\text{Planck}\right)^2}{\left(\sigma_\text{rel}\times\Omega h^2|_\text{DM,\,total}\right)^2}\,,
\end{equation}
where we assume that the dominant uncertainty originates from the theoretical prediction of the
relic density, in particular from the uncertainty on the annihilation cross section. We estimate 
$\sigma_\text{rel}=10\%$ (\emph{cf.}~\emph{e.g.}~\cite{Arroyo:2016wal} for a discussion of 
the one-loop corrections to the Higgs portal type annihilation and \cite{Banerjee:2016vrp} for a relevant discussion within the IDM). The corresponding log-likelihood 
finally reads
\begin{equation}
-2 \log \mathcal{L}_\Omega   =     \chi_\Omega^2   
+ 2 \log (\sigma_{\text{rel}}\,\Omega h^2|_\text{DM,\,total}),
\end{equation}
up to an irrelevant constant.

%- - - - - - - - - - - - - - - - - - - - - - - - - - - - - - - - - - - - - -
\subsubsection{Direct detection}  \label{sec:LUX}
%- - - - - - - - - - - - - - - - - - - - - - - - - - - - - - - - - - - - - -

In the IDM at tree-level spin-independent WIMP-nucleon scattering
arises from $Z$- and $h$-exchange~\cite{Barbieri:2006dq}. 
The former is, however, only significant for extremely small mass splittings
between $\Hn$ and $\An$, which are not in the focus of this study.
The cross section for WIMP-nucleon scattering via $h$-exchange reads
\begin{equation} 
\label{eq:sigmaSI}
\sigma_\text{SI} = \frac{\lambda_L^2 f_\text{N}^2}{4 \pi}\frac{\mu_\text{r}^2 m_{\rm N}^2}{m_{\hn}^4 m_{\Hn}^2}\,,
\end{equation}
where $f_\text{N} \sim 0.30$~\cite{Cline:2013gha} denotes the strength of the effective 
Higgs-nucleon interaction, and $\mu_{\rm r} = m_{\rm N} m_{\Hn}/(m_{\rm N}+m_{\Hn})$ 
is the DM-nucleon reduced mass. 
In this study we compute the total spin-independent scattering cross section at tree-level 
(including an effective vertex for the Higgs interaction with gluons) with \textsc{micrOMEGAs}~\cite{Belanger:2014vza}.

We take into account the most recent constraints from LUX~\cite{Akerib:2016vxi}. In order to
estimate the respective log-likelihood we utilize the program 
\textsc{LUXCalc}~\cite{Savage:2015xta}. However, the current version of \textsc{LUXCalc} 
is based on the results from LUX 2013~\cite{Akerib:2013tjd}. In order to account for the 
LUX 2016 sensitivity we proceed as follows. We first determine
the (mass-dependent) gain factor in the sensitivity between the LUX 2013 and LUX 2016. 
Assuming that the improvement in the sensitivity 
is well described by a simple gain in the exposure we then rescale the signal by this factor 
and compute the log-likelihood with \textsc{LUXCalc}~\cite{Savage:2015xta}.
Computing the $p$-value in this way allows us to reproduce the limits from LUX 2016
with a relative uncertainty at a per-cent level in the mass region of interest.

It should be noted that as \eqref{eq:sigmaSI} is proportional to $\lambda_L^2$, for very small $\lambda_L$ electroweak
corrections induced at 1-loop can be important and can eventually dominate the scattering
cross section~\cite{Klasen:2013btp,Abe:2015rja}. However, the magnitude of the
electroweak corrections is independent of $\lambda_L$ and below $\sim2\times10^{-47}\text{cm}^2$
for the mass regions considered in this study~\cite{Klasen:2013btp}. Hence, they are well 
below the current sensitivity of LUX~\cite{Akerib:2016vxi} and can be neglected for the
computation of the respective likelihood. We will discuss their importance for direct detection
projections in section~\ref{sec:prospects}.

%- - - - - - - - - - - - - - - - - - - - - - - - - - - - - - - - - - - - - -
\subsubsection{Indirect detection constraints: dwarf spheroidal galaxies} \label{sec:constraints_indirect}
%- - - - - - - - - - - - - - - - - - - - - - - - - - - - - - - - - - - - - -

In the low mass region of the IDM where annihilation occurs predominantly into $b\bar b$ and/or $W^+W^-$ pairs,
the most stringent limits on the velocity-averaged annihilation cross section, $\langle \sigma v \rangle$,
arise from gamma-ray observations of dSphs. We use the results of the recent analysis of the 
Fermi-LAT data~\cite{Fermi-LAT:2016uux}.
\\
\\
The predicted  $E^2 \times$ flux in an energy bin between $E_{\min}$ and $E_{\max}$ is
\begin{equation}
E^2 \frac{\D \phi}{\D E}=\frac{1}{4\pi} \frac{\langle\sigma v\rangle R^2}{2 m_\text{DM}^2} \int_{E_{\min}}^{E_{\max}} \D E_\gamma \,E_\gamma\frac{\D N_\gamma}{\D E_\gamma}\times \int_\text{ROI} \D \Omega \int_\text{l.o.s}\D s \, \rho_\text{DM}^2  \,,
\end{equation}
where $\D N_\gamma/\D E_\gamma$ is the differential photon spectrum per annihilation
and $m_\text{DM}$ is the mass of the DM particle. The integral of the DM density,
$\rho_\text{DM}$, over the region of interest (ROI) and the line of sight (l.o.s) is the $J$-factor, $J_i$, of the 
considered dwarf.

In order to take into account constraints from dSphs in our fit
we use tabulated likelihoods for individual dwarfs as a function of the energy 
flux in the considered 24 energy bins provided in \cite{Fermi-LAT:2016uux}. The total
log-likelihood is obtained by summing over the log-likelihood contributions of 
the individual dwarfs \cite{Fermi-LAT:2016uux,Ahnen:2016qkx}. 
We take into account the seven dwarfs with the largest confirmed $J$-factors 
\cite{Geringer-Sameth:2014yza} per default, see first seven dwarfs 
listed in table~\ref{tab:dwarfs}.

We compute the prediction for the binned energy flux by using the tabulated
gamma-ray spectra for individual annihilation channels obtained in~\cite{Cuoco:2016jqt}, which we combine 
according to their relative strength as calculated with \textsc{micrOMEGAs}~\cite{Belanger:2014vza}.
These channels include the 3-body final states $WW^*$, $ZZ^*$, where the virtual vector boson creates a pair of fermions.
The uncertainty of the $J$-factors is taken into account by profiling over the
$J_i$ (for each dwarf) according to its uncertainty provided in table~\ref{tab:dwarfs}, \emph{i.e.}, 
for each sampled point we tabulate \emph{on-the-fly} the likelihood of the considered dwarfs 
as a function of $J_i$ and take the corresponding value that provides the maximum likelihood.

\begin{table}[h]
\begin{center}
\renewcommand{\arraystretch}{1.2}
\begin{tabular}{c | c } 
dwarf  & $\log_{10}( J_\text{meas})\;[ \,\log_{10} (\text{GeV}^2\,\text{cm}^{-5})\,]$ \\ \hline
Coma Berenices & $19.0\pm0.4 $ \\
Draco & $18.8\pm0.1$ \\
Sculptor & $18.5\pm0.1$ \\
Segue 1 & $19.4\pm0.3$ \\
Ursa Major II & $19.4\pm0.4$ \\
Ursa Minor & $18.9\pm0.2$ \\ 
Reticulum II & $18.9\pm0.6$\\
\hline
Tucana III & $19.3\pm0.6$ \\
Tucana IV &   $18.7\pm0.6$
\end{tabular}
\renewcommand{\arraystretch}{1}
\end{center}
\caption{$J$-factors and their uncertainties for the dwarf spheroidal galaxies
considered in this study. For the first seven dwarfs we use the measured values
from \cite{Geringer-Sameth:2014yza}. For Tucana III and Tucana IV $J$-factors are
estimated, see text for details. 
}
\label{tab:dwarfs}
\end{table}

In four of the recently discovered dSph targets, slight excesses 
(each $\sim2\sigma$ local) have been found: specifically in
Reticulum II, Tucana III, Tucana IV and Indus II~\cite{Geringer-Sameth:2015lua,Li:2015kag,Fermi-LAT:2016uux} 
(but see also~\cite{Drlica-Wagner:2015xua}).
For the latter three targets the dynamical masses have not yet been spectroscopically 
measured and hence these targets are at present not confirmed as DM-dominated dSphs. However, in order to
illustrate the impact of the respective observation on the fit we consider the
case of additionally including the log-likelihood of Tucana III and Tucana IV in the fit. 
For these targets we use the distance-based predictions for the $J$-factors
with an estimated error of 0.6 dex~\cite{Fermi-LAT:2016uux}.

It should be noted that the uncertainties in the $J$-factors used above might be
underestimated, comparing \emph{e.g.}~\cite{Geringer-Sameth:2014yza,Martinez:2013els}
and, in particular, following the discussion in~\cite{Bonnivard:2015xpq}.
However, these uncertainties have only a minor impact on our results. Omitting the
likelihood contribution from the faint dwarfs Coma Berenices, Ursa Major II and Segue 1 
(that exhibit the largest uncertainties~\cite{Bonnivard:2015xpq} among the considered ones)
does not qualitatively change our results.

Finally, we note that additional constraints could stem from the Fermi-LAT searches for gamma-ray lines at the Galactic center~\cite{Ackermann:2015lka}. Within the IDM line signatures have been studied in~\cite{Gustafsson:2007pc,Garcia-Cely:2015khw}. Although the loop-suppression of the production cross section for two monochromatic photons is typically compensated by the higher sensitivity in searches for spectral lines, we do not expect these searches to provide constraints significantly stronger than the ones for a continuous photon spectrum in dwarf spheroidal galaxies considered above.\footnote{%
For instance, for the benchmark point IV in~\cite{Gustafsson:2007pc} ($m_h = 120\,\text{GeV}$, $m_\Hn=70\,\text{GeV}$ and $\lamL\simeq -0.07$)
the predicted line signal, $\langle\sigma v\rangle_{\gamma\gamma}=7.6\times10^{-30}\,\text{cm}^3\text{s}^{-1}$, falls below the upper limit, $\langle\sigma v\rangle_{\gamma\gamma}^\text{UL}=5.2\times10^{29}\,\text{cm}^3\text{s}^{-1}$~\cite{Ackermann:2015lka}, by almost an order of magnitude while the cross section into $b\bar b$ alone, $\langle\sigma v\rangle_{b\bar b}=1.6\times10^{26}\,\text{cm}^3\text{s}^{-1}$, is already relatively close to the respective limit from dwarfs, $\langle\sigma v\rangle_{b\bar b}^\text{UL}=2.6\times10^{26}\,\text{cm}^3\text{s}^{-1}$~\cite{Fermi-LAT:2016uux}. Furthermore, the above quoted limit from line searches is derived for the most aggressive choice regarding the dark matter density profile considered in~\cite{Ackermann:2015lka}, \emph{i.e.}~a generalized Navarro-Frenk-White profile with an inner slope of $\gamma=1.3$, which is, however, compatible with the GCE\@. For choices of the dark matter density profile that are more cored the limit becomes even weaker. Note also that in~\cite{Gustafsson:2007pc} the 3-body final state contribution to the continuous gamma spectrum have not been taken into account.}
However, a full assessment of the importance of line searches in the considered parameter space of the IDM falls beyond the scope of this work.

%- - - - - - - - - - - - - - - - - - - - - - - - - - - - - - - - - - - - - -
\subsubsection{The Galactic center excess} \label{sec:constraints_indirect}
%- - - - - - - - - - - - - - - - - - - - - - - - - - - - - - - - - - - - - -

%-   -   -   -   -   -   -   -   -   -   -   -   -   -   -   -   -   -   -
\subsubsection*{The Fermi-LAT observation}\label{sec:Jfactor}
%-   -   -   -   -   -   -   -   -   -   -   -   -   -   -   -   -   -   -

Over the last few years several groups have reported a Galactic bulge emission 
component in the Fermi-LAT gamma-ray data~\cite{Goodenough:2009gk,Hooper:2010mq,Hooper:2011ti,Abazajian:2012pn,Hooper:2013rwa,Gordon:2013vta,Abazajian:2014fta,Daylan:2014rsa,Zhou:2014lva,Calore:2014xka,Agrawal:2014oha,Calore:2014nla,Huang:2015rlu,TheFermi-LAT:2015kwa,Karwin:2016tsw} -- the GCE\@.
Extending beyond~10$^\circ$ away from the Galactic plane,  
the GCE appears compatible with a spherical morphology and 
a steep cuspy radial profile~\cite{Daylan:2014rsa,Calore:2014xka}. 
In the $E^2 \times$ flux representation the inferred energy spectrum peaks at a few GeV.
Intriguingly, the excess is compatible with a signal from
DM annihilation. In particular it favors an annihilation cross section close to the thermal one,
$\langle\sigma v\rangle\sim 10^{-26}\,\text{cm}^3\text{s}^{-1}$, which could point towards explanations in terms of thermal WIMPs (but see also~\cite{TheFermi-LAT:2017vmf}).
Besides interpretations in terms of DM, various astrophysical explanations of the excess have been 
proposed~\cite{Petrovic:2014uda,Petrovic:2014xra,Yuan:2014rca,Cholis:2015dea}.
Using statistical methods to indicate whether the photon-count distributions of the excess are compatible with 
a smooth component, exhibiting Poissonian clustering properties, evidence
for an extended gamma-ray point source population have been 
found~\cite{Lee:2014mza,Lee:2015fea,Bartels:2015aea,McDermott:2015ydv} 
disfavoring the DM interpretation of the GCE, see also~\cite{Clark:2016mbb}.
However, it remains difficult to control the systematic uncertainties in point source analyses
that could arise due to mismodeling of the data~\cite{Horiuchi:2016zwu}. Hence, it is premature
to draw a definite conclusion about the origin of the GCE\@. 

In this study we consider the possibility (results presented in section~\ref{sec:fitGCE}) 
that the excess could be entirely due to WIMP 
annihilation. We use the results of the analysis performed in~\cite{Calore:2014xka}, which provided
the inferred energy spectrum along with an error
covariance matrix that includes an estimate of systematic uncertainties related to the Galactic 
foreground emission.

%-   -   -   -   -   -   -   -   -   -   -   -   -   -   -   -   -   -   -
\subsubsection*{Dark matter density profile and uncertainties}\label{sec:Jfactor}
%-   -   -   -   -   -   -   -   -   -   -   -   -   -   -   -   -   -   -

The DM density in the inner part of the Milky Way is subject to large 
uncertainties affecting the observed gamma-ray flux which is reflected by
an uncertainty in the involved $J$-factor
\begin{equation}\label{eq:jfactor}
J_{40^\circ} =  \int\limits_{\text{ROI}}\!\D\Omega \! \int\limits_{\textrm{l.o.s}}\!\D s \,\rho^2_\text{DM}\,,
\end{equation}
where the ROI is a $40^\circ \times 40^\circ$
region centered on the Galactic center with a stripe of $\pm 2^\circ$ masked along 
the Galactic plane~\cite{Calore:2014xka}.
The DM spatial distribution in the Milky Way has been evaluated using dynamical data, for example, in~\cite{Pato:2015dua,Iocco:2016itg} and the ensuing uncertainties
on the $J$-factor stem dominantly from the poor knowledge of the inner slope of the DM halo profile (for a recent study of the impact of these uncertainties on the DM-induced
gamma-ray flux \textit{cf. e.g.}~\cite{Benito:2016kyp}). However, the GCE cannot be reproduced unless specific assumptions are made concerning the DM spatial distribution. 
Concretely, using a generalized Navarro-Frenk-White (NFW) profile~\cite{Navarro:1995iw} with parameters compatible with the measured shape of the GCE, \emph{i.e.}~an inner slope of $\gamma=1.2\pm0.08$~\cite{Calore:2014xka},\footnote{For a further discussion of the morphology of the GCE see \emph{e.g.}~\cite{Calore:2014nla} and references therein.}
as well as recent measurements of the scale radius and the scale density~\cite{Nesti:2013uwa},
based on a Monte Carlo procedure 
it was shown in~\cite{Cuoco:2016jqt} that the $J$-factor is
\begin{equation}
\label{eq:Jnom}
\log \left(\frac{J_{40^\circ}}{\,\textrm{GeV}^2 \textrm{cm}^{-5}}\right) = 23.25\pm 0.43\,.
\end{equation}
Note that since the authors of \cite{Calore:2014xka} normalize the GCE flux dividing by the angular size of the analyzed region,
we divide $J_{40^\circ}$ by the corresponding solid angle, $\Delta\Omega=0.43$ sr.

%-   -   -   -   -   -   -   -   -   -   -   -   -   -   -   -   -   -   -
\subsubsection*{Likelihood for the GCE signal}\label{sec:gcefit}
%-   -   -   -   -   -   -   -   -   -   -   -   -   -   -   -   -   -   -

We compute the $\chi^2$ contribution for the GCE -- including the contribution from $J_{40^\circ}$ -- by 
\begin{equation}
\label{eq:chi2GCE}
\chi_\text{GCE}^2 = \sum_{i,j} (d_i - t_i ) \left( \Sigma_{ij} + \delta_{ij} (\sigma_{\text{rel}}\ t_i)^2\right)^{-1}(d_j -  t_j) 
 + \frac{(\log J_{40^\circ} - \log J_{40^\circ\!,\,\text{nom}} )^2}{(\sigJ)^2},
\end{equation}
where $d_i$ is  the  GCE measured flux in energy bin $i$ from \cite{Calore:2014xka} and
$t_i$ is the respective model prediction, which depends on the model parameters, $R$ and $J_{40^\circ}$.
$\Sigma_{ij}$ is the covariance matrix given in~\cite{Calore:2014xka} and $\log J_{40^\circ\!,\,\text{nom}}$ and 
$\sigJ$ are the nominal values of the (logarithmic) $J$-factor and its uncertainty, respectively, as given in \eqref{eq:Jnom}.
We compute the predicted flux $t_i$ by combining the tabulated
gamma-ray spectra for individual annihilation channels(including the 3-body final states $WW^*$, $ZZ^*$) obtained in~\cite{Cuoco:2016jqt}
weighted by their relative contributions as computed by \textsc{micrOMEGAs}~\cite{Belanger:2014vza}.
In addition to the covariance matrix which includes statistical and systematic errors of the observed signal
we include $\delta_{ij} (\sigma_{\text{rel}}\ t_i)^2$ representing  a diagonal error equal to a fraction $\sigma_{\text{rel}}$
of the model prediction. We choose $\sigma_{\text{rel}}$ = 10\% as discussed in~\cite{Caron:2015wda,Bertone:2015tza,Cuoco:2016jqt}.
Up to an irrelevant constant factor, the resulting log-likelihood is
\begin{equation}
\label{eq:logLGCE}
-2 \log \mathcal{L}_\text{GCE}   =     \chi_\text{GCE}^2   + \log | \Sigma_{ij} + \delta_{ij} (\sigma_{\text{rel}}\ t_i)^2 |,
\end{equation}
where $| \Sigma_{ij} + \delta_{ij} (\sigma_{\text{rel}}\ t_i)^2 |$ is the determinant of the covariance matrix.

%------------------------------------------------------------------------
\subsection{Non-dark matter observables}
%------------------------------------------------------------------------

%------------------------------------------------------------------------
\subsubsection{Unitarity, perturbativity and vacuum stability} \label{sec:theobounds}
%------------------------------------------------------------------------

Besides experimental constraints, it is important to impose a minimal set of theoretical requirements which ensure that the results we will obtain are reliable and physically meaningful. To this goal we use the \textsc{2HDMC} code~\cite{Eriksson:2009ws}. First, we demand that all couplings be perturbative, which amounts to a condition $\left| \lambda_i \right| < 4\pi$ for all couplings appearing in \eqref{Eq:TreePotential}. Secondly, in any perturbative calculation the scattering matrix should be unitary order-by-order in perturbation theory. Failure of such a condition is typically associated with the development of strong dynamics which, again, renders a perturbative treatment unreliable. The tree-level scalar and vector scattering amplitudes are required to be smaller than $16\pi$, \emph{i.e.}~we allow that the unitarity limit be saturated already at tree-level. Lastly, the electroweak vacuum should be sufficiently long-lived. Here we impose the condition already implemented in \textsc{2HDMC} which simply requires that the vacuum be completely stable. We note that metastable vacua in the IDM have been studied in \cite{Khan:2015ipa,Swiezewska:2015paa}. 

In our scanning procedure, parameter space points failing at least one of these requirements are immediately discarded, \emph{i.e.}~these theoretical requirements are imposed as ``hard'' constraints. In practice, we assign them a large enough negative log-likelihood pushing them well outside the $4\sigma$ region around the best-fit points.

%------------------------------------------------------------------------
\subsubsection{Higgs invisible width} \label{sec:BRinv}
%------------------------------------------------------------------------

When $m_\HO < m_{\hn}/2$, the decay $h \to \HO\HO$ is allowed and contributes to the invisible decay width of the Higgs boson as
\begin{equation}
\Gamma\left( \hn \rightarrow \Hn\Hn \right) = \frac{\lambda_L^2 v^2}{8\pi m_{\hn}} \sqrt{1 - \frac{4 m_{\Hn}^2}{m_{h}^2}}
\end{equation}
In this region of parameter space, the coupling of a pair of $\Hn$ to $\hn$ is constrained by LHC measurements which set an upper bound, ${\rm BR}_{\rm inv} \lesssim 0.23$~\cite{Aad:2015pla}. For the numerical analysis
we use the log-likelihood function for ${\rm BR}_{\rm inv}$ provided in~\cite{Aad:2015pla}.

%------------------------------------------------------------------------
\subsubsection{Electroweak precision observables} \label{sec:EWprec}
%------------------------------------------------------------------------

The new states which are present in the IDM can contribute to the $S$, $T$ and $U$ 
oblique parameters \cite{Altarelli:1990zd,Peskin:1990zt}. Deviations from the Standard 
Model expectations in $U$ are negligible \cite{Goudelis:2013uca}. Hence we assume 
the latter to be zero and consider only $S$ and $T$. We compute their $\chi^2$ contribution
through
\begin{equation}
\chi^2_{ST} = \vec v^T {\cal C}^{-1}\vec v
\end{equation}
where $\vec v^T \equiv (S-\hat S, T - \hat T)$ and ${\cal C}$ is the covariance matrix. For $U = 0$, the electroweak fit performed in \cite{Baak:2014ora} gives the values
\begin{equation}
\hat S = 0.06\,,\quad \hat T = 0.097
\end{equation}
for a reference Higgs mass of $125$ GeV, while the covariance matrix is given by
\begin{equation}
{\cal C}= 
\begin{pmatrix}
0.0085 & 0.0063\\
0.0063 & 0.0057
\end{pmatrix}\,.
\end{equation}
The contributions of the new scalar states to $S$ and $T$ are computed with \textsc{2HDMC}~\cite{Eriksson:2009ws}.

%------------------------------------------------------------------------
\subsubsection{LEP-II bounds on the masses of the heavy ${\cal{Z}}_2$-odd states} \label{sec:LEPII}
%------------------------------------------------------------------------

The masses of the heavier ${\cal{Z}}_2$-odd states can be constrained by translating the corresponding mass bounds from searches for charginos and neutralinos at LEP-II. The former were recast in \cite{Pierce:2007ut}, yielding a rough bound on the mass of the charged states:
\begin{equation} \label{eq:mHpmLEP}
m_{\Hp} \gtrsim m_W \ .
\end{equation}
Limits on the mass of the heavier neutral state (which, we remind, in our case we take to be $\An$) were extracted in \cite{Lundstrom:2008ai} and amount to 
\begin{equation}\label{eq:mA0LEP}
m_{\An} \gtrsim 100~\mathrm{GeV} \ .
\end{equation}
In the subsequent analysis, we will simply restrict our scanning regions to 
$m_{\An},m_{\Hp} \gtrsim 100~\mathrm{GeV}$  in order for these limits to be (conservatively) satisfied, 
without including them in our global fit analysis. 

%------------------------------------------------------------------------
\subsection{Scan settings}
%------------------------------------------------------------------------

In order to perform the global fit we use \textsc{MultiNest}~\cite{Feroz:2008xx,Feroz:2013hea},
which allows an efficient scan of the parameter space under investigation. 
The considered parameters and respective scan ranges are summarized in table~\ref{tab:scanparams}.

Although \textsc{MultiNest} is particularly suited for Bayesian analyses, in this 
work we will solely adopt a frequentist interpretation. This is possible provided 
that the posterior, and hence the resulting likelihood, has been explored in sufficient
detail. This approach has two advantages. On the one hand, the derived 
constraints are not dependent on the priors chosen to explore the parameters.
On the other hand, we can combine the output of different \textsc{MultiNest} scans,
which allows an efficient use of the generated chains and provides the possibility 
to specifically improve on the coverage of the parameter space.
However, in this approach the density of points, which in the Bayesian interpretation has
a precise meaning (namely, it traces the posterior distribution), does not have any 
physical relevance anymore. 

To ensure that the likelihood is sampled in enough detail, we run multiple \textsc{MultiNest} scans with 
high-accuracy settings, using between 600 and 3000 live points, a tolerance between $\texttt{tol}=$ 0.1--0.001, 
and an enlargement factor between $\texttt{efr}=0.3$--$0.6$ in order
to achieve a good exploration of the tails of the distribution. 

For the resulting fits we perform marginalization over parameters with the profile likelihood 
method \cite{Rolke:2004mj} and draw contours at a certain confidence level following the 
expectation of a (two-dimensional) $\chi^2$ distribution.

\begin{table}[h]
\begin{center}
\renewcommand{\arraystretch}{1.2}
\begin{tabular}{c | c} 
parameter & range  \\ \hline
$m_{\Hn}$ & $[45;1000]\,\text{GeV}$\\
$m_{\An}$ & $[100^*;1100]\,\text{GeV}$\\
$m_{\Hp}$ & $[100^*;1100]\,\text{GeV}$\\
$\lamL$  & $[-4\pi\,;\,4\pi]$\\ 
$\lam_2$  & $[10^{-6}\,;\,4\pi]$\\ \hline
$\logJ$& $[-4\sigJ ; 4\sigJ]$\\
$R$ & $[10^{-3};1]$
\end{tabular}
\renewcommand{\arraystretch}{1}
\end{center}
\caption{Parameters of the fit and the corresponding allowed ranges. 
\!${}^*$Additionally we require $m_{\An},m_{\Hp}>m_{\Hn}$.}
\label{tab:scanparams}
\end{table}

%===================================================================
\section{Results and discussion} \label{sec:results} 
%===================================================================

We now proceed to discuss our main results. As a first step, we update on the status of the IDM by performing a global fit including all 
constraints and observables described in section~\ref{sec:constraints} except the GCE spectrum and the two unconfirmed dwarfs  
described in section~\ref{sec:constraints_indirect} (Tucana III and IV). Subsequently, we include the GCE as well as the new dwarfs and check whether 
the IDM can provide an explanation for the GCE whilst satisfying all other constraints.

%------------------------------------------------------------------------
\subsection{Global dark matter fit}\label{sec:noGCEres}
%------------------------------------------------------------------------

\begin{figure}[h!]
\centering
\setlength{\unitlength}{1\textwidth}
\begin{picture}(0.95,0.94)
 \put(0.0,0.03){ 
  \put(0.03,-0.01){\includegraphics[width=0.88\textwidth]{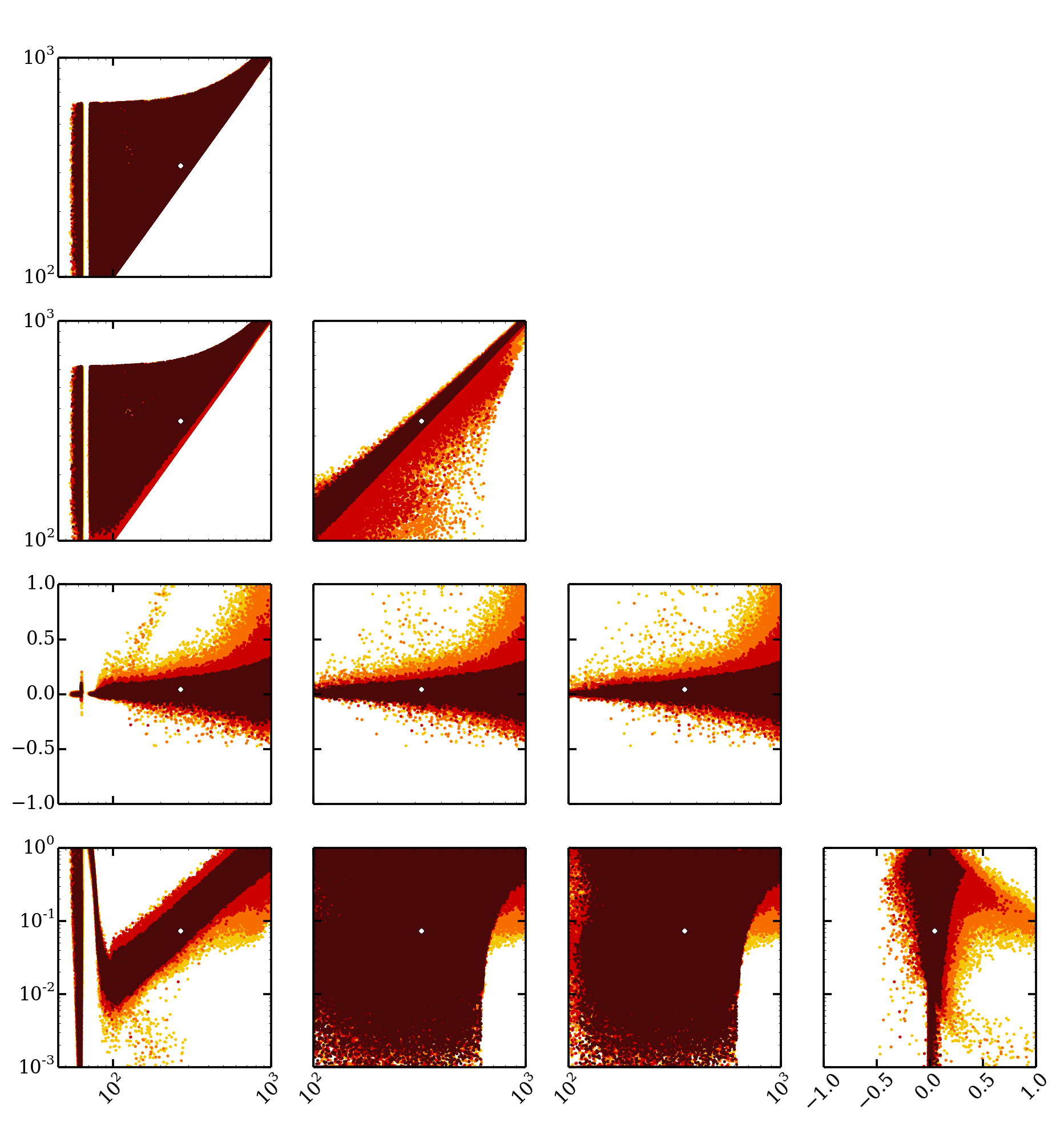}}
  \put(0.02,0.743){\rotatebox{90}{\scriptsize $m_{\An}\,[\text{GeV}]$}}
  \put(0.02,0.523){\rotatebox{90}{\scriptsize $m_{\Hp}\,[\text{GeV}]$}}
  \put(0.01,0.343){\rotatebox{90}{\scriptsize $\lambda_L$}}
  \put(0.01,0.127){\rotatebox{90}{\scriptsize $R$}}
  \put(0.127,-0.015){\scriptsize $m_{H^0}\,[\text{GeV}]$}
  \put(0.336,-0.015){\scriptsize $m_{\An}\,[\text{GeV}]$}
  \put(0.542,-0.015){\scriptsize $m_{\Hp}\,[\text{GeV}]$}
  \put(0.796,-0.015){\scriptsize $\lambda_L$}
  }
\end{picture}
\caption{Global fit results for the IDM free parameters $m_{\Hn}$, $m_\An$, $m_\Hp$, $\lambda_L$ and the DM fraction $R$.
The brown, red, orange and yellow points lie within 1-, 2-, 3- and $4\sigma$ away from the best-fit point (denoted by a white dot), respectively. Here we take into account
the log-likelihood contributions from all observables described in section \ref{sec:constraints}, except the GCE spectrum and unconfirmed dwarfs.}
\label{fig:triangle_wo_gce}
\end{figure}
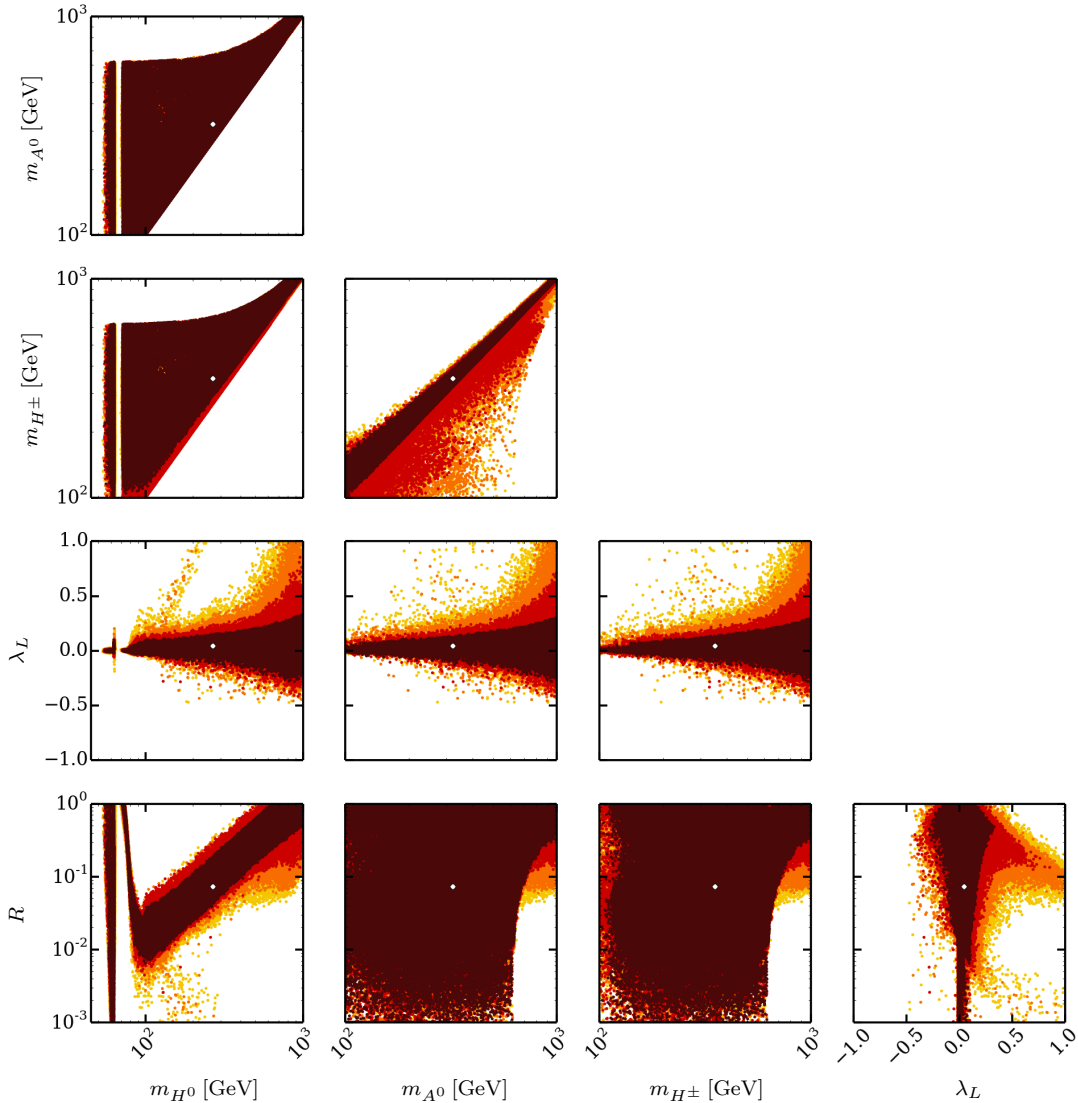

The results of our global fit are presented in figure \ref{fig:triangle_wo_gce}, where we show projections of the parameter space 
defined in table~\ref{tab:scanparams} onto 2-dimensional planes of all combinations of the involved parameters. As  $\logJ$ only concerns the GCE it is not included in the plot.
Furthermore, we do not show $\lambda_2$ as it is virtually featureless, since the considered observables do not have any dependence on $\lambda_2$ at leading order in perturbation 
theory. We highlight 1-, 2-, 3- and 4$\sigma$ regions of the log-likelihood (brown, red, orange and yellow points respectively) around the best-fit point 
(represented by a white dot). The Higgs boson mass is fixed at $m_{\hn} = 125.09$ GeV. 

A first observation that can be made by inspecting figure \ref{fig:triangle_wo_gce} is that the masses of the ${\cal{Z}}_2$-odd scalars cannot vary arbitrarily with respect to
each other. Concretely, the mass splitting between the lightest ($\Hn$ in our case) and the next-to-lightest ${\cal{Z}}_2$-odd particle cannot exceed a few hundred GeV: it can 
reach maximally $\sim 500$ GeV for $m_{\Hn} = 100$ GeV and decreases to $\sim 300$ GeV for $m_{\Hn} = 500$ GeV. This behavior is a consequence of the perturbativity requirement: 
from the relations \eqref{eq:physmasses} we observe that $m_{\Hn}^2 - m_{\An}^2 = \lambda_5 v^2$, implying that indeed large mass splittings between the two scalars can drive the $\lambda_5$
coupling to non-perturbative values. At the same time, the mass difference between $\An$ and $\Hp$ has to be small due to the constraints
from the $S$, $T$ parameters described in section \ref{sec:EWprec}: generically, contributions to $S$ and $T$ are due to the breaking of the custodial symmetry of the scalar potential
\eqref{Eq:TreePotential} which is induced by the $\lambda_{4,5}$ terms. Assuming all parameters to be real, the symmetry can be restored if and only if 
$\lambda_4 = \lambda_5$ \cite{Haber:2010bw}, whereas deviations from this condition amount to contributions to the oblique parameters. At the same time, from the relations
\eqref{eq:physmasses} we see that $m_{\An}^2 - m_{\Hp}^2 = (\lambda_4 - \lambda_5) v^2/2$. This, in turn, implies that the contributions to $S$ and $T$ increase 
as the mass splitting between the two states becomes large.

Secondly, from figure \ref{fig:triangle_wo_gce} we see that the IDM can account for the entire DM abundance in the Universe ($R \sim 1$) in three distinct $m_{\Hn}$ regions: 
one centered around half the Higgs mass (the so-called ``funnel region''), one around 72 GeV and, finally, for relatively large $m_{\Hn} \gtrsim 500$ GeV. The general reasons for
this behavior have been analyzed in the literature \cite{LopezHonorez:2006gr,Hambye:2009pw,Honorez:2010re,LopezHonorez:2010tb,Goudelis:2013uca}: ignoring, for the moment, direct 
detection constraints, the IDM can reproduce the observed DM abundance in the Universe in three $m_{\Hn}$ regions. The first corresponds to $m_{\Hn} < m_W$, where annihilation proceeds
through the Higgs portal - type process $\Hn\Hn \rightarrow \hn \rightarrow f\bar{f}/VV^*$ (with $V = W^{\pm} /Z$) as well as through direct annihilation via the point-like
$\Hn$-$\Hn$-$V$-$V$ vertex. Annihilation into virtual gauge bosons increases in importance as the corresponding kinematic threshold is approached from below. 
Besides, the LEP constraints on the heavier ${\cal{Z}}_2$-odd scalar masses described in section \ref{sec:LEPII} exclude the possibility of coannihilation in this mass range.
Once $m_{\Hn}$ becomes larger than $m_W$, annihilation into gauge bosons becomes dominant. If fact, it is \textit{too} efficient, so destructive interference must occur between the
Higgs portal - like diagram and the one involving the four-vertex, which for a Higgs mass of $\sim 125$ GeV can happen for negative (positive) values of $\lambda_L$ if $m_{\Hn} > m_{\hn}/2$ ($m_{\Hn} < m_{\hn}/2$). Above roughly 120 GeV, this interference cannot be efficient enough so that DM becomes necessarily underabundant (this
upper value depends on the Higgs boson mass \cite{LopezHonorez:2010tb}).
%=====================
%    \                                           |
%      \                                         |
%        \                                       |
%                                      \         |
%                                        \       |
%                                          \     |
%=====================

The Imposition of direct detection constraints wipes out most of this parameter space, since the DM-nucleon scattering cross section is proportional to $\lambda_L^2$. 
The only regions surviving are those characterized by very small values of $\lambda_L$, which 
correspond either to the Higgs funnel region or to the regime where annihilation into pairs of virtual gauge bosons becomes efficient enough (but not too efficient so that cancellations
are needed) without requiring large values of $\lambda_L$.

In the high-mass regime, additional effects come into play. As described in detail in \cite{Hambye:2009pw}, once $m_{\Hn} \gtrsim 500$ GeV the destructive interference between the annihilation diagram involving the quartic $\Hn$-$\Hn$-$V$-$V$ coupling and $t$-channel diagrams involving the heavier ${\cal{Z}}_2$-odd particles can become efficient enough so as to bring the $\Hn$ self-annihilation cross section down to acceptable levels. The cancellation becomes more exact the smaller the mass splitting between the inert states, and in practice the condition $\left| m_{\Hn} - m_{\An/\Hp} \right| \lesssim 10$ GeV must be satisfied otherwise $\left\langle \sigma v \right\rangle$ is too large (this is actually also dictated by unitarity arguments). The predicted DM abundance then depends on the interplay of this cancellation mechanism with the contributions from $s$-channel exchange of a Higgs boson and coannihilation. If the cancellation is exact, large enough values of $\lambda_L$ (of ${\cal{O}}(0.3)$ \cite{Goudelis:2013uca}) are needed in order to saturate the Planck bound. Besides, since in this regime DM is relatively heavy, it is only poorly constrained by direct/indirect detection and/or collider searches.

Dropping the $R \sim 1$ requirement, \textit{i.e.}~going to scenarios in which the IDM only accounts for a fraction of the total DM content of the Universe, allows for more freedom in both $m_{\Hn}$ and $\lambda_L$. This effect is particularly pronounced in the $\Hn$ mass range between roughly 75 and 500 GeV, where DM naturally tends to be underabundant due to the efficiency of the 
direct $\Hn$-$\Hn$-$V$-$V$ coupling. In Higgs-portal models, it is typically the same coupling that controls the annihilation and the WIMP-nucleon scattering cross section. 
This implies that in such scenarios underabundant WIMP dark matter, $R<1$, does not amount to weaker direct detection bounds, since to first order both the relic density and direct detection constraints induce the same relation between the Higgs portal coupling and the WIMP fraction, $\lambda\propto R^{-1/2}$.
In the IDM this is true only for regions where Higgs-mediated annihilation is clearly dominant. Above $\sim72\,$GeV the direct $\Hn$-$\Hn$-$V$-$V$ interaction dominates the annihilation
while it does not enter the direct detection cross section at leading order. This introduces more freedom in
$m_{\Hn}$ and $\lambda_L$ for $R<1$. Note that this is true only for the current direct detection 
constraints that have not yet reached sensitivity to the loop-induced electroweak corrections~\cite{Klasen:2013btp,Abe:2015rja}. For future direct detection experiments this situation can change, \emph{cf.} the discussion in section~\ref{sec:prospects}.

With direct detection constraints being largely inefficient for DM masses above $\sim 75$ GeV, the behaviour of the allowed values of the relic abundance as a function of $m_{\Hn}$ is largely determined by the interference pattern between the direct $\Hn$-$\Hn$-$V$-$V$ coupling and $t/u$-channel diagrams involving the heavier ${\cal{Z}}_2$-odd particles \cite{Hambye:2009pw}. In particular, the largest allowed values of $R$ correspond to the smallest mass splittings between the Inert Doublet scalars. This might appear counter-intuitive, since from a Boltzmann suppression standpoint this is the regime where coannihilation effects should become the most relevant. However, this is also the regime in which the quartic couplings in the scalar potential \eqref{Eq:TreePotential} vanish. In this limit, the cancellation between four-vertex interactions and $t/u$-channel diagrams involving the heavier scalars becomes maximal, with a similar remark also applying to coannihilation processes. Turning on the quartic couplings, 
\emph{i.e.}~increasing the mass splitting between the Inert Doublet scalars, can only \textit{increase} the total $\Hn$ (co-)annihilation cross section and, hence, decrease the predicted DM abundance. We thus see that in the underabundant IDM region the upper limit on $R$ corresponds to quasi-degenerate Inert Scalars, whereas the lower one to large mass splittings within the Inert Doublet (as well as to large values of $\lambda_L$). 

%------------------------------------------------------------------------
\subsection{Fitting a possible signal from dark matter annihilation}
 \label{sec:fitGCE}
%------------------------------------------------------------------------

\begin{figure}[h!]
\centering
\setlength{\unitlength}{1\textwidth}
\begin{picture}(1.03,1.03)
 \put(0.0,0.0){ % big one
  \put(0.03,0.01){\includegraphics[width=0.97\textwidth]{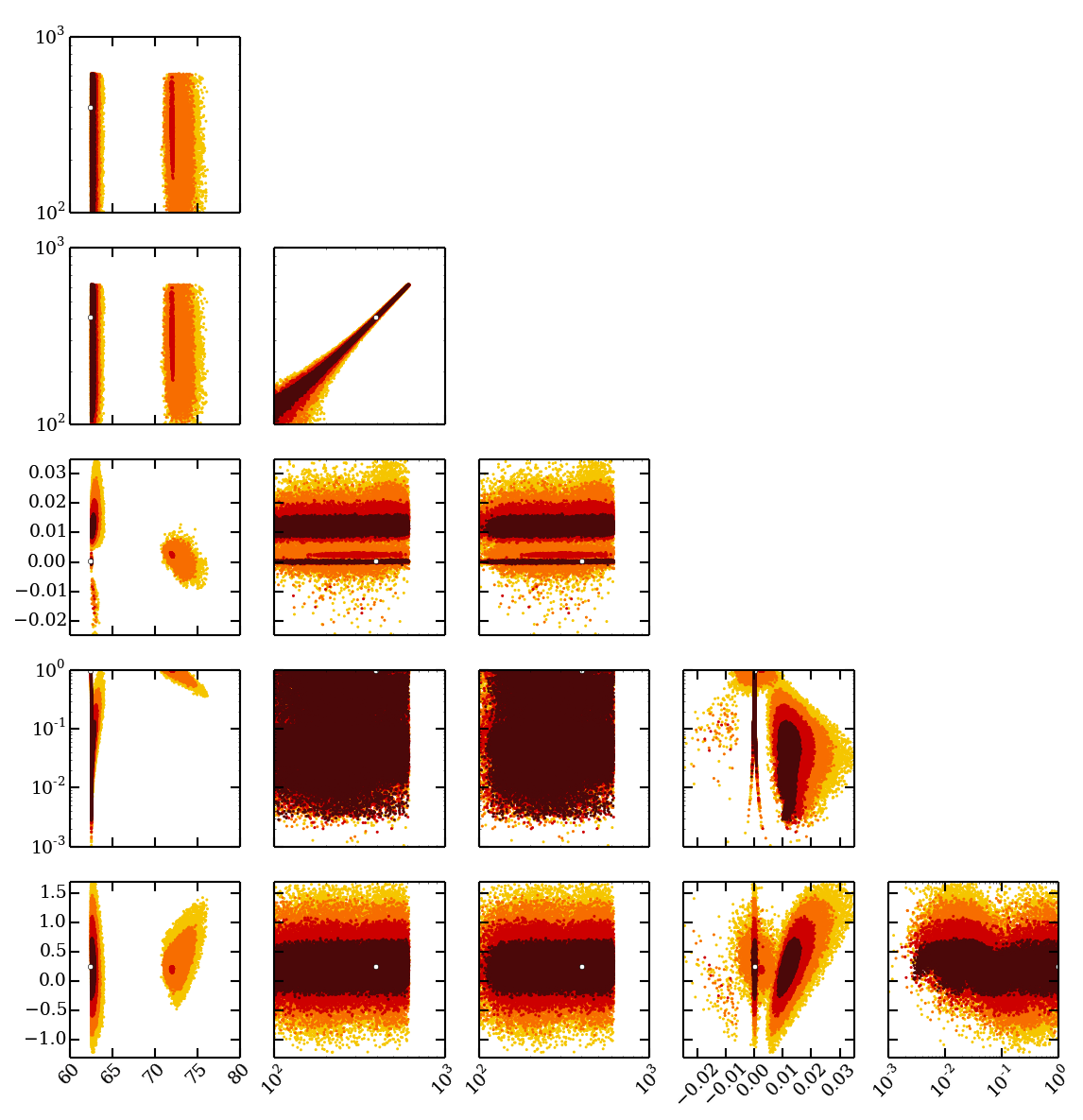}}
  \put(0.32,0.88){ (a)}
  \put(0.02,0.855){\rotatebox{90}{\scriptsize $m_{\An}\,[\text{GeV}]$}}
  \put(0.02,0.666){\rotatebox{90}{\scriptsize $m_{\Hp}\,[\text{GeV}]$}}
  \put(0.02,0.517){\rotatebox{90}{\scriptsize $\lambda_L$}}
  \put(0.02,0.328){\rotatebox{90}{\scriptsize $R$}}
  \put(0.02,0.097){\rotatebox{90}{\scriptsize $\logJ$}}
  \put(0.125,0.005){\scriptsize $m_{H^0}\,[\text{GeV}]$}
  \put(0.3,0.005){\scriptsize $m_{\An}\,[\text{GeV}]$}
  \put(0.48,0.005){\scriptsize $m_{\Hp}\,[\text{GeV}]$}
  \put(0.71,0.005){\scriptsize $\lambda_L$}
  \put(0.89,0.005){\scriptsize $R$}
  }
\put(0.6045,0.603){ % zoom
\includegraphics[width=0.371\textwidth]{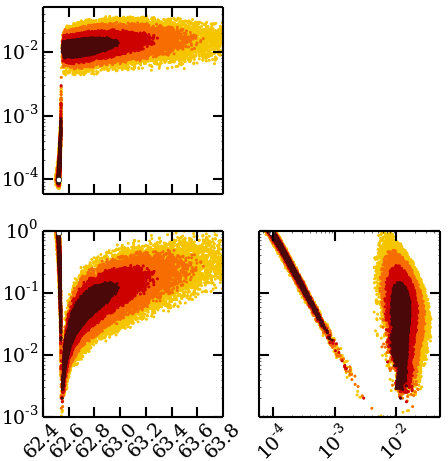}
\put(-0.405,0.107){\rotatebox{90}{\scriptsize $R$}}
\put(-0.405,0.278){\rotatebox{90}{\scriptsize $\lambda_L$}}
\put(-0.276,-0.025){\scriptsize $m_{H^0}$}
\put(-0.095,-0.025){\scriptsize $\lambda_L$}
}
  \put(0.88,0.88){ (b)}

\end{picture}
\caption{Same as figure \ref{fig:triangle_wo_gce} but including the GCE contribution to the log-likelihood and including the $\logJ$
parameter in the fitting procedure.
The large triangle (a) shows the correlation between all parameters. The reduced triangle (b)
focusses on the Higgs resonance region (region 1) with positive $\lambda_L$ (plotted logarithmically).
}
\label{fig:triangle_w_gce}
\end{figure}
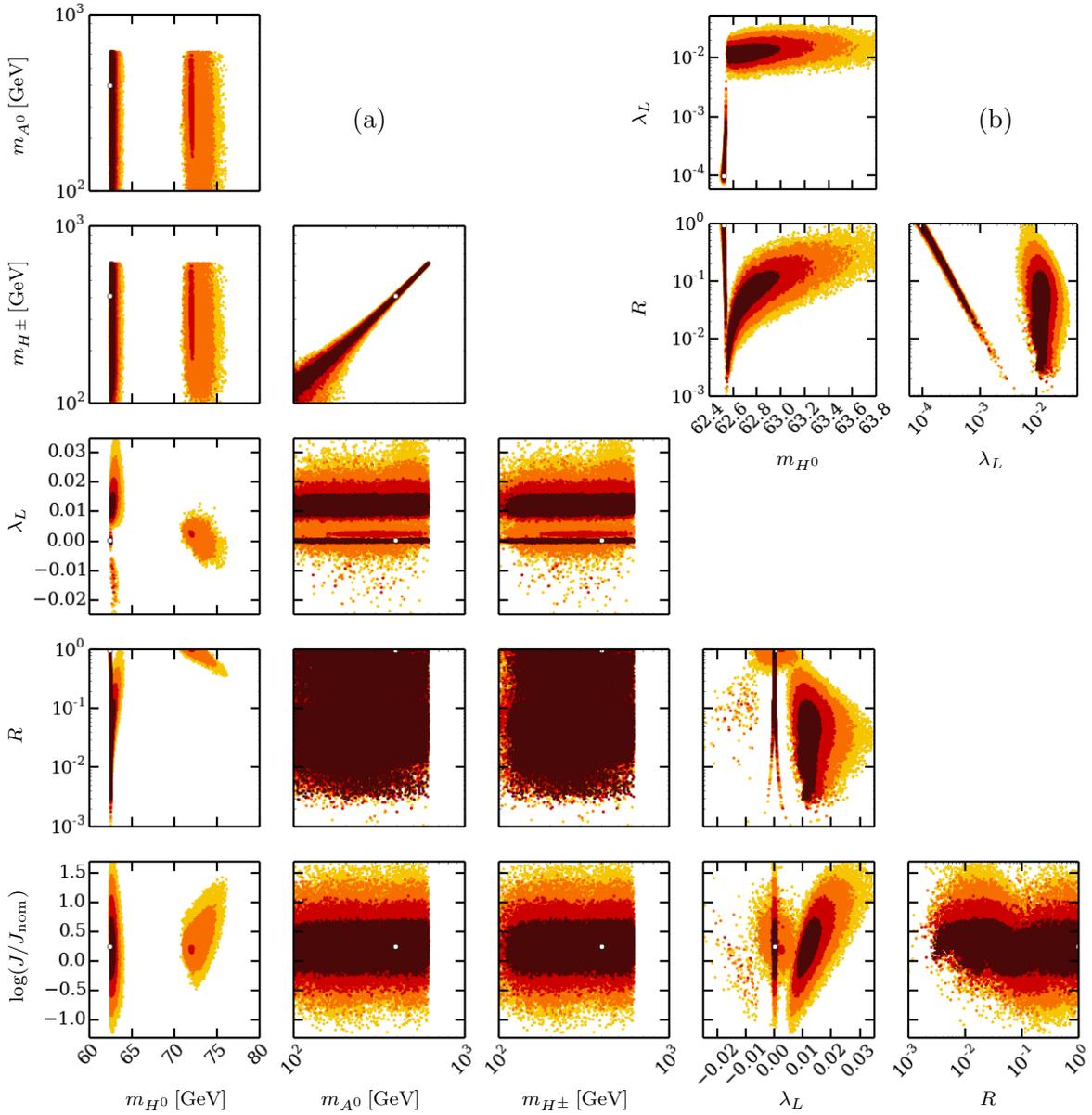

\begin{figure}[h]
\centering
\setlength{\unitlength}{1\textwidth}
\begin{picture}(0.6,0.325)
 \put(0.0,0.0){ 
  \put(-0.01,-0.01){\includegraphics[width=0.6\textwidth]{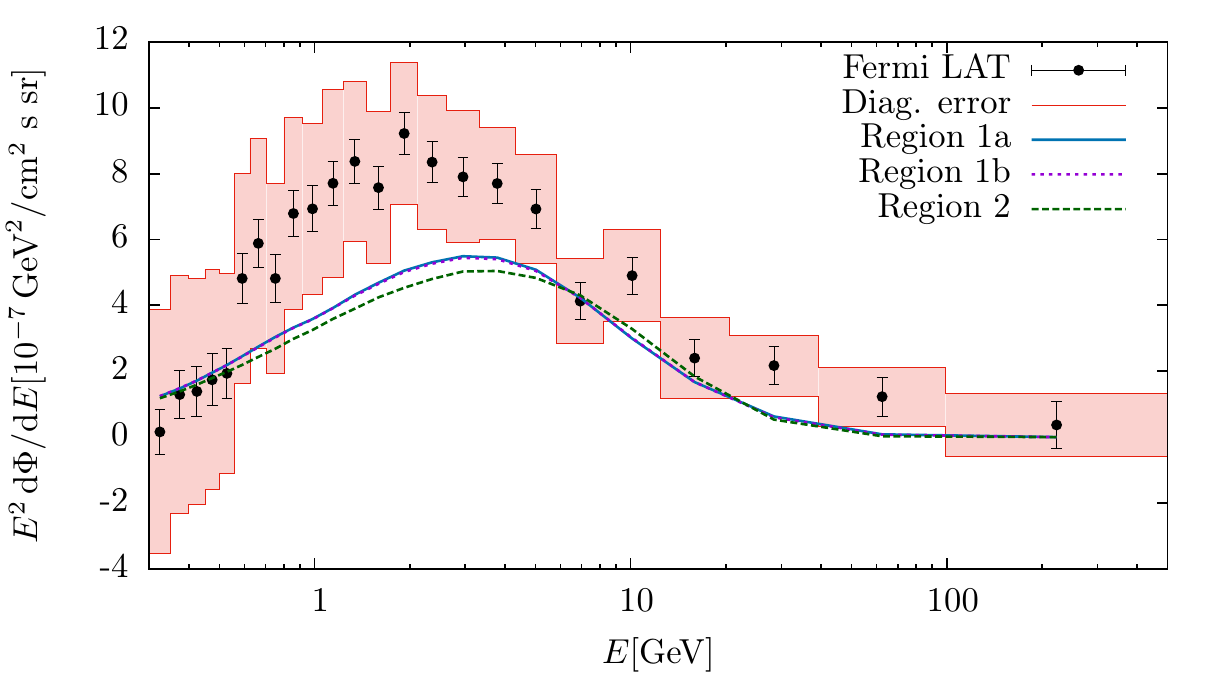}}
  }
\end{picture}
\caption{Comparison of the gamma-ray spectra predicted by the IDM best-fit points with the GCE spectrum.
}
\label{fig:spectra}
\end{figure}

We now examine whether the IDM can accommodate the GCE as a signal from DM annihilation. In addition to the observables considered in section~\ref{sec:noGCEres} we include the GCE likelihood in our global fit as described in section \ref{sec:constraints_indirect}. The result of the fit is shown in figure~\ref{fig:triangle_w_gce}
(again, we omit $\lambda_2$ since it does not enter any of the observables we consider). We find two distinct regions in parameter space in which the IDM can explain the
GCE\@. The respective best-fit points are summarized in table~\ref{tab:best-fit} and the corresponding spectra are shown in figure~\ref{fig:spectra}. 

%=====================
%    \                                           |
%      \                                         |
%        \                                       |
%                                      \         |
%                                        \       |
%                                          \     |
%=====================

\begin{table}
\begin{centering}
\renewcommand{\arraystretch}{1.15}
\begin{tabular}{c | c c c}
 & region 1a & region 1b & region 2 \\ \hline\hline
$m_{H^0}[\textrm{GeV}]$ & $62.522^{+ 0.017}_{- 0.004}$ & $62.61^{+ 0.23}_{- 0.06}$ & $71.95^{+ 1.18}_{- 0.48}$ \\
$\lambda_{L}$ & $0.00009^{+0.00039}_{-0.00001}$ & $0.011^{+0.003}_{-0.002}$ & $|\lambda_L|<0.005$ \\
$R$ & $0.99^{+0.01}_{-0.97}$ & $0.02^{+0.07}_{-0.02}$ & $1.00^{+0.00}_{-0.19}$ \\
$\log\left( J / J_{nom} \right)$ & $0.25^{+0.31}_{-0.38}$ & $0.24^{+0.33}_{-0.37}$ & $0.19^{+0.25}_{-0.16}$ \\
$\chi^2_{\textrm{GCE}}$ & $26.9$ & $27.3$ & $33.6$ \\
$p(\chi^2_{\textrm{GCE}})$ & $0.22$ & $0.20$ & $0.054$ \\
$p(\textrm{LUX})$ & $1$ & $0.79$ & $0.67$ \\
$p(\textrm{dSph})$ & $0.33$ & $0.33$ & $0.34$ \\
$p(\textrm{rel.D.})$ & $0.87$ & $0.92$ & $0.99$ \\
$p(\textrm{ST})$ & $0.68$ & $0.67$ & $0.68$ \\
dom. channel & $b\bar b \,(81\%)$ & $b\bar b \,(67\%)$ & $WW^* \,(89\%)$ \\
\end{tabular}
\protect\caption{\label{tab:best-fit}Parameters describing the GCE best-fit regions appearing in figure \ref{fig:triangle_w_gce}. The regions 1a and 1b correspond to the Higgs
funnel with  $m_{\Hn} \approx m_{\hn}/2$. Region 2 refers to the case where DM annihilates predominantly into virtual gauge boson pairs. The last line indicates the dominant annihilation channel and its relative contribution to annihilation today.
}
\par\end{centering}
\end{table}

%-   -   -   -   -   -   -   -   -   -   -   -   -   -   -   -   -   -   -
\subsubsection*{Higgs funnel region}
%-   -   -   -   -   -   -   -   -   -   -   -   -   -   -   -   -   -   -

The first region lies close to the Higgs funnel, where DM annihilation 
proceeds predominantly via $s$-channel Higgs exchange near the resonance,	
$m_{\Hn} \simeq m_{\hn}/2$. In figure~\ref{fig:triangle_w_gce}{b} we zoom into this part of the parameter space, restricting $\lambda_L$ to positive values in order to
allow for a logarithmic scaling (we will comment on the asymmetry regarding positive and negative
$\lambda_L$ below). 
Upon closer inspection we can see that this region splits up into two subregions.
In the first one (hereafter referred to as region 1a), $m_{\Hn}$ is restricted to an extremely narrow range between 62.5\,GeV and 62.55\,GeV
(\emph{i.e.}~just below $m_{\hn}/2$)
and $\lambda_L$ to small values $10^{-4}\lesssim\lambda_L\lesssim10^{-2}$. In this case the IDM can also account
for the entire DM abundance in the Universe, $R\simeq1$. In fact, large $R$ is slightly preferred as we will explain further below.
In the second subregion (region 1b), $m_{\Hn}$ is slightly larger, $\lambda_L\sim10^{-2}$ and the fit favors 
$R\lesssim0.2$. 

This structure is mainly driven by the interplay of two observables, namely the DM relic density and the GCE itself. In the vicinity of the resonance 
the annihilation cross section exhibits a large velocity dependence. Hence, a small variation in $m_{\Hn}$ can alter the ratio 
$\sv_\text{today}/\sv_\text{freeze-out}$ by orders of magnitude. In particular, for $m_{\Hn}$ just
above $m_{\hn}/2$, this ratio can become larger than $10^2$. Moreover, the gamma-ray flux
and the predicted relic DM density scale differently with the faction $R$: 
the relic density scales as $\Omega h^2|_\text{DM,\,total} \propto 1/( R \left\langle \sigma v \right\rangle_{\rm freeze-out} )$ whereas the gamma-ray flux is proportional to $\left\langle \sigma v \right\rangle_{\rm today} \times R^2$.
Now, the likelihood function is minimised for $\Omega h^2|_\text{DM,\,total} \sim \Omega h^2|_\text{Planck}$, which in turn implies $\left\langle \sigma v \right\rangle_{\rm freeze-out} \times R \sim \left\langle \sigma v \right\rangle_{\rm thermal}$. 
At the same time, in order to reproduce the GCE (which can be explained by an annihilation cross section of the order of the thermal one) a flux corresponding to $\sv_{\rm thermal}\sim\left\langle \sigma v \right\rangle_{\rm today} \times R^2 $ is required. Hence, our fit globally prefers parameter space regions in which
\begin{equation}
\frac{\left\langle \sigma v \right\rangle_{\rm today}}{\left\langle \sigma v \right\rangle_{\rm freeze-out}} \times R \sim {\cal{O}}(1) \, .
\end{equation}
As $R \leq 1$, this condition can only be satisfied if $\left\langle \sigma v \right\rangle_{\rm today} \ge \left\langle \sigma v \right\rangle_{\rm freeze-out}$, a situation occurring for $m_{\Hn} \gtrsim m_{\hn}/2$. 

In the best-fit region 1a 
$m_{\Hn}$ is finely tuned to a value very close to (but slightly smaller than) $m_{\hn}/2$. The minimal mass in this region corresponds roughly to 
$\sv_\text{today}/\sv_\text{freeze-out} \simeq 1$ and, hence, $R=1$. 
As $m_{\Hn}$ approaches $m_{\hn}/2$ from below, $\sv_\text{today}/\sv_\text{freeze-out}$ increases extremely rapidly for very small variations of $m_{\Hn}$ and, hence, the gamma-ray flux remains large enough despite the reduction in the overall DM density as explained before.
Region 1b sets in for slightly larger masses just above $m_{\hn}/2$, where $\sv_\text{today}/\sv_\text{freeze-out}$ is again large but gradually starts to decrease. However, this region does not extent up to masses for which $\sv_\text{today}/\sv_\text{freeze-out}\simeq1$ due to an interplay of two aspects. On the one hand, for increasing mass the shape of the gamma-ray spectrum (mainly due to annihilation into $b\bar b$) tends to yield a worse fit of the GCE. On the other hand, the larger $\lambda_L$ values required start to be in tension with limits from direct detection.

This behavior is similar to the one found for the singlet scalar 
Higgs portal model in~\cite{Cuoco:2016jqt}. However, there are some differences worth commenting upon. In the singlet scalar 
model, around the Higgs funnel region DM annihilation only occurs via $s$-channel Higgs exchange. Hence, the 
relative contribution to annihilation does not depend on the coupling between the DM particle and the Higgs.
In the IDM the additional four-vertex interaction $\Hn$-$\Hn$-$V$-$V$ interferes with
$\Hn\Hn\to h\to VV^{(*)}$, which introduces some additional features. 

%=====================
%    \                                           |
%      \                                         |
%        \                                       |
%                                      \         |
%                                        \       |
%                                          \     |
%=====================

In region 1a $\lambda_L$
becomes as small as $10^{-4}$ for $R\simeq1$. For such small values of $\lambda_L$ the
contributions of the four-vertex and Higgs exchange diagrams are of the same order and, 
hence, there is a strong destructive (constructive) interference for positive (negative) $\lambda_L$
in the annihilation cross section into vector bosons.\footnote{We remind that the interference is destructive (constructive) for positive (negative) $\lambda_L$ when $s<(m_{\hn}/2)^2$ which, in the zero-velocity limit, amounts to $m_{\Hn}<m_{\hn}/2$. The situation is inversed for $s>(m_{\hn}/2)^2$ \cite{LopezHonorez:2010tb}.} This leads to a large variation of the $WW^*$
contribution to the total gamma-ray spectrum, ranging from a negligible fraction for small positive $\lambda_L$ to around 50\% 
for small negative $\lambda_L$. Since the $b\bar{b}$ annihilation spectrum provides
a better fit of the GCE than the $WW^*$ one, the fit tends to favor
a small positive $\lambda_L\simeq10^{-4}$ coupling, where the $WW^*$ contribution is below the percent
level. This is the origin of the asymmetry in $\lambda_L$ and furthermore explains the preference
for large $R$ values in region 1a -- due to the strong correlation between $\lambda_L$ and $R$.
For larger values of $\lambda_L$ the Higgs-exchange diagram dominates and the $WW^*$ contribution 
becomes comparable to the one in the singlet scalar model.

In region 1b the situation is reversed and we obtain a smaller $WW^*$ contribution for negative values
of $\lambda_L$. Although, following the same line of reasoning, the GCE should now favor negative $\lambda_L$ values,
the overall likelihood is actually maximized for positive ones. This is caused by the DM relic density contribution to the total likelihood. 
Due to the strong velocity-dependence the various annihilation 
channels behave very different during freeze-out and at present times and in this case
interference effects become important for larger couplings, thus giving rise to the required annihilation
cross section for positive $\lambda_L$ only. This causes a strong preference for positive $\lambda_L$
in region 1b, despite the slightly worse GCE fit (given the larger $WW^*$ contribution). These remarks also explain the
overall preference for region 1a in the fit, contrary to the case of the singlet scalar Higgs portal 
model analysed in~\cite{Cuoco:2016jqt}.

%-   -   -   -   -   -   -   -   -   -   -   -   -   -   -   -   -   -   -
\subsubsection*{Region around 72\,GeV}
%-   -   -   -   -   -   -   -   -   -   -   -   -   -   -   -   -   -   -

The second mass region where the IDM can explain the GCE lies around $m_{\Hn}=72$ GeV.
In this case the fit to the GCE is considerably worse than in regions 1,
$\chi^2/\text{dof}=33.6/22$.
However, it globally remains at acceptable levels (within 2$\sigma$ from the best-fit point of region 1a),
which is partly driven by the other observables considered, \emph{cf.}~table~\ref{tab:best-fit}.
Moreover, this region does not require a large degree of fine-tuning of $\lambda_L$ and $m_{\Hn}$
and strongly favors that the IDM accommodates the full observed DM abundance ($R=1$) through sufficiently efficient annihilation of $\Hn$ pairs into virtual gauge bosons via the quartic $\Hn$-$\Hn$-$V$-$V$ interaction.
The corresponding coupling is a pure gauge coupling, completely independent of $\lambda_L$, and the latter can, hence, be tuned to sufficiently small values in order to evade current direct detection constraints. As the DM mass increases, and $m_{\Hn}$ approaches the $WW$ threshold, this process gradually becomes too efficient which would rather imply $R < 1$, \textit{cf.}~the discussion in section~\ref{sec:noGCEres}. However, the spectrum for annihilation into $WW^*$ provides a considerably worse fit to the GCE for masses above $m_{\Hn}=72\,$GeV. (In fact, for pure $WW^*$ annihilation a DM mass around $55\,$GeV would fit best.) Note that $\sv_\text{today}$ is somewhat smaller than $\sv_\text{freeze-out}$ due to the kinematic suppression away from the $WW$ threshold for small
velocities. As a consequence the IDM tends to undershoot the
required flux for fitting the GCE which is reflected in the preference for positive $\logJ$. The gamma-ray spectra resulting from the best-fit points of regions 1a, 1b and 2 are shown for comparison in figure \ref{fig:spectra}, along with the corresponding Fermi-LAT measurements.

\bigskip

\begin{figure}[h!]
\centering
\setlength{\unitlength}{1\textwidth}
\begin{picture}(0.55,0.525)
\put(0.0,0.03){
  \put(0.03,-0.01){\includegraphics[width=0.52\textwidth]{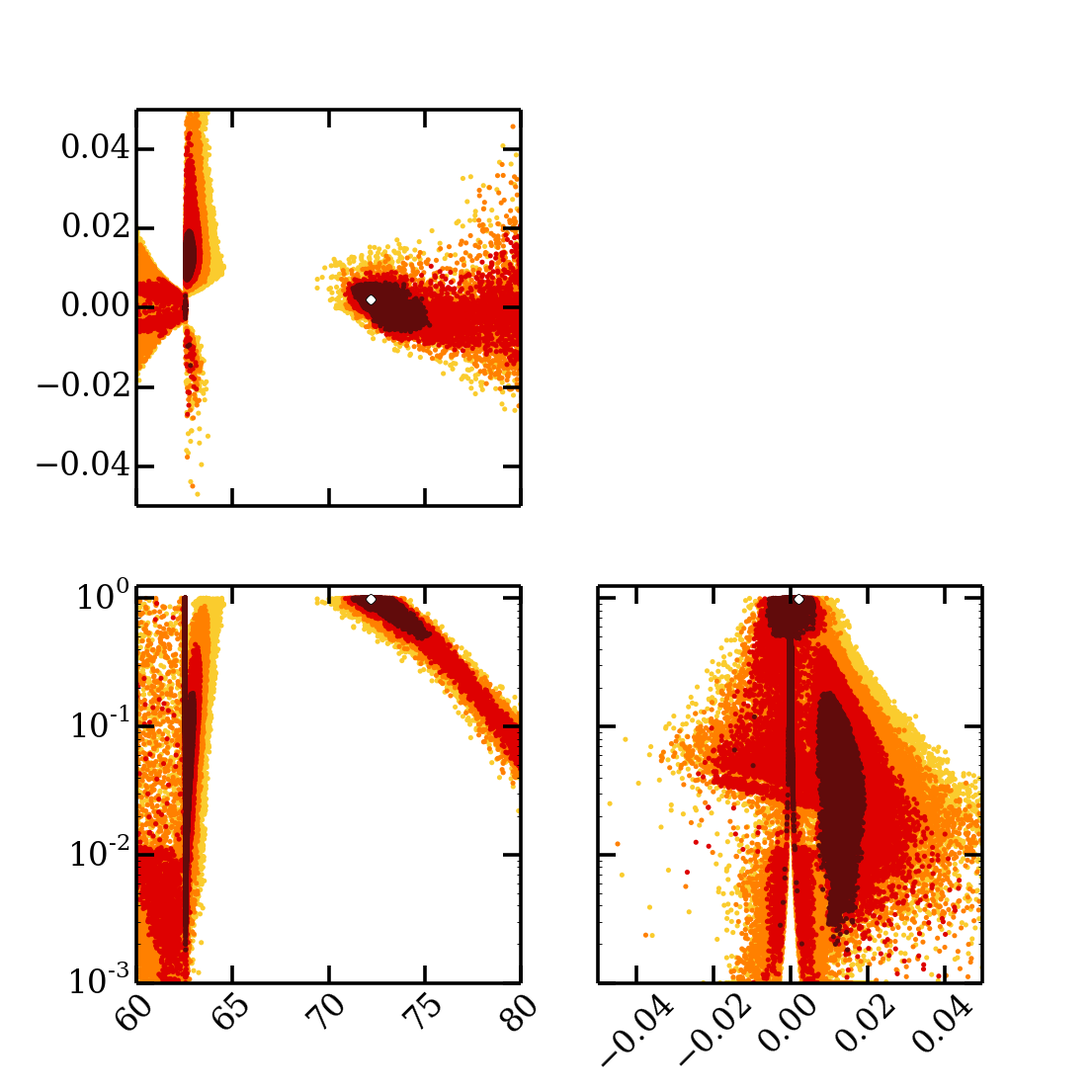}}
\put(0.02,0.35){\rotatebox{90}{\scriptsize $\lambda_L$}}
\put(0.02,0.123){\rotatebox{90}{\scriptsize $R$}}
\put(0.144,-0.02){\scriptsize $m_{H^0}\,[\text{GeV}]$}
\put(0.397,-0.02){\scriptsize $\lambda_L$}
}
\end{picture}
\caption{As in figure \ref{fig:triangle_w_gce}b but including unconfirmed dwarfs instead of the GCE.
}
\label{fig:triangle_w_gce_w_dwarfs}
\end{figure}
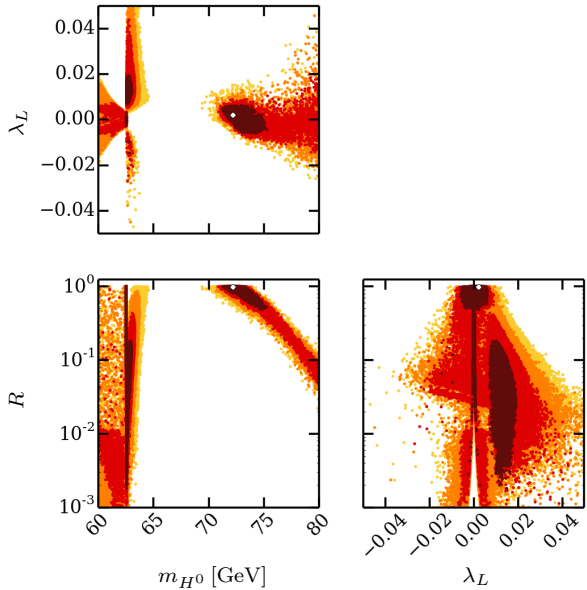

In our fit so far we only considered confirmed dSph targets, for which the $J$-factor has been measured.
In four of the recently discovered dSph targets, slight excesses 
(each $\sim2\sigma$ local) have been found: Reticulum II, Tucana III, 
Tucana IV and Indus II~\cite{Li:2015kag,Fermi-LAT:2016uux}. It is therefor interesting to see
in how far these excesses are compatible with the DM explanation of the GCE\@. To this end,
we include the likelihood contribution of Tucana III and IV in our fit.
The result is shown in figure~\ref{fig:triangle_w_gce_w_dwarfs} where we omit the GCE likelihood
contribution in order to allow for a comparison. We indeed find the same two regions as before. 
Interestingly, the overall best-fit point now lies in region 2.
However, given that the excess is mild the preference for the best-fit point with respect to, \emph{e.g.}, the high mass region, 
$m_{\Hn} \gtrsim 600\,\text{GeV}$ (that does not fit the excess), is only at the level of 1--2$\sigma$. 

%=====================
%    \                                           |
%      \                                         |
%        \                                       |
%                                      \         |
%                                        \       |
%                                          \     |
%=====================

Let us also note that recently a hint for a possible DM annihilation signal has also been 
found in the AMS-02 antiproton data, pointing to a DM mass around 60--80\,GeV and
a cross section around $3\times10^{-26}\text{cm}^2\text{s}^{-1}$~\cite{Cuoco:2016eej}.
Although for a given annihilation channel this excess favors a somewhat larger cross section 
than the GCE, these two observations are still compatible with each other taking into account the 
uncertainties on the local DM density~\cite{Cuoco:2017rxb}, in particular for the annihilation 
channels $b\bar{b}$ and $WW$ which are important in the IDM\@.

%------------------------------------------------------------------------
\subsection{Future prospects for direct, indirect detection and the LHC}\label{sec:prospects}
%------------------------------------------------------------------------

Finally, let us briefly discuss the future prospects for direct detection, indirect detection and
collider searches in the IDM, focusing in particular on the GCE best-fit regions found in section~\ref{sec:fitGCE}.

%-   -   -   -   -   -   -   -   -   -   -   -   -   -   -   -   -   -   -
\subsubsection*{Direct detection}
%-   -   -   -   -   -   -   -   -   -   -   -   -   -   -   -   -   -   -

In the near future the sensitivity of direct detection experiments is expected to improve even further with the advent of new experiments. 
The expected 90\% CL exclusion
limit on the spin-independent DM-nucleon scattering cross section for a DM mass
between 62 and 72\,GeV is estimated at $(1.7\!-\!1.8)\times 10^{-47}\,\text{cm}^2$ for
Xenon1T~\cite{Aprile:2015uzo}, $(2.4\!-\!2.5)\times 10^{-48}\,\text{cm}^2$ for
LZ~\cite{Akerib:2015cja} and 
$(2.8\!-\!3.0)\times 10^{-49}\,\text{cm}^2$ for Darwin~\cite{Aalbers:2016jon}.
The latter improves by up to three orders of magnitude upon the sensitivity of 
LUX 2016~\cite{Akerib:2016vxi} which we took into account in our fitting procedure and which 
provides a limit in the vicinity of $(2.2\!-\!2.3)\times 10^{-46}\,\text{cm}^2$ for the same mass range.
The three GCE best-fit regions found in section~\ref{sec:fitGCE} are characterized by small $\lambda_L$
and, hence, a small DM-nucleon scattering cross section at tree-level. Especially for regions 1a and 2, our findings show that 
the tree-level scattering cross section is too small to be challenged by upcoming experiments.
However, as mentioned in section~\ref{sec:LUX}, electroweak radiative corrections provide a 
contribution to the DM-nucleon scattering that is independent of $\lambda_L$, and which
can actually dominate over the tree-level contribution. For the best-fit points in region 1a and 2 the radiative corrections are of the
order of $1.5\times 10^{-48}\,\text{cm}^2$~\cite{Klasen:2013btp} which brings them slighly outside
the projected sensitivity of LZ but well within the reach of Darwin. 

Note that for points with $R<1$ within the same mass range, the electroweak corrections lose importance
with respect to the tree-level contribution. Such points correspond to larger values of $\lambda_L$, for which the decrease in the signal 
due to the small value of $R$ is roughly compensated by the larger $\lambda_L$-induced scattering cross section, which is not true for the electroweak contribution.

Besides, upcoming direct detection experiments can also provide handles for the high-mass region of the IDM.
As discussed in section~\ref{sec:noGCEres}, in the range between 100 and 600 GeV where 
the IDM tends to provide too little DM, the largest $R$ values are obtained
for small mass splittings between $\Hn$ and the heavier inert doublet partners, as well as
for small values of $\lambda_L$. This is true in particular for the points with the smallest
masses that allow for $R=1$ around 600 GeV. For these points, the contribution
from electroweak corrections is likely to dominate the tree-level WIMP-nucleon scattering cross section since $\lambda_L$ can be 
made sufficiently small so as to render the Higgs-mediated DM-nucleon scattering negligible.
The corresponding cross section (in the limit $\lambda_L=0$) is about
$1.4\times 10^{-47}\,\text{cm}^2$~\cite{Klasen:2013btp} for a DM mass of 600 GeV.
This value is just within the projected reach of LZ~\cite{Akerib:2015cja}, which will, hence, start to push 
the lower mass limit of the IDM high-mass region (assuming $R=1$).

%-   -   -   -   -   -   -   -   -   -   -   -   -   -   -   -   -   -   -
\subsubsection*{Indirect detection}
%-   -   -   -   -   -   -   -   -   -   -   -   -   -   -   -   -   -   -

\begin{figure}[t]
\centering
\setlength{\unitlength}{1\textwidth}
\begin{picture}(0.95,0.375)
 \put(0.0,0.0){ %left 
   \put(0.0,0.03){\includegraphics[width=0.45\textwidth]{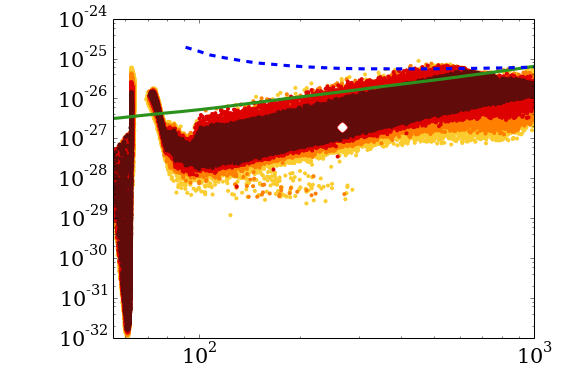}}
  \put(0.01,0.145){\rotatebox{90}{\scriptsize $\sv\times R^2$}}
  \put(0.214,0.01){\scriptsize $m_{H^0}\,[\text{GeV}]$}
  \put(0.13,0.259){\rotatebox{6.2}{\tiny {\color{darkgreen} Fermi-LAT}}}
  \put(0.131,0.246){\rotatebox{6.2}{\tiny {\color{darkgreen} (dSphs 15y)}}}
  \put(0.192,0.289){\rotatebox{-6}{\tiny {\color{blue} CTA }}}
  \put(0.23,0.285){\rotatebox{-2}{\tiny {\color{blue} (GC 100\,hr)}}}
  }
 \put(0.47,0.0){ %right
   \put(0.0,0.03){\includegraphics[width=0.45\textwidth]{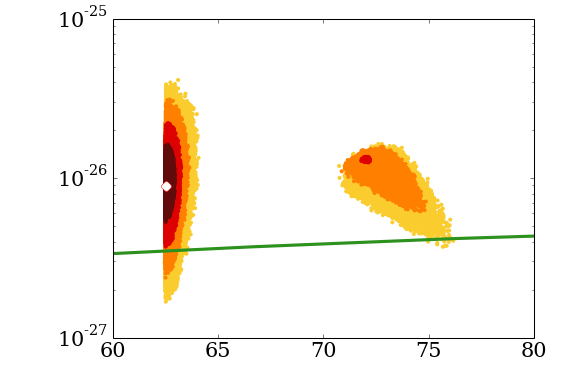}}
  \put(0.01,0.145){\rotatebox{90}{\scriptsize $\sv\times R^2$}}
  \put(0.214,0.01){\scriptsize $m_{H^0}\,[\text{GeV}]$}
  \put(0.18,0.157){\rotatebox{2.7}{\tiny {\color{darkgreen} Fermi-LAT}}}
  \put(0.181,0.14){\rotatebox{2.7}{\tiny {\color{darkgreen} (dSphs 15y)}}}
  }
\end{picture}
\caption{
Fit without (left) and with (right) the GCE in the $m_{H^0}$-$\sv R^2$ plane (color code as in figure~\ref{fig:triangle_wo_gce}) superimposed by indirect detection projections. The Fermi-LAT dSphs projections (green lines) are derived for annihilation into $b\bar b$
which is the main annihilation cross section around 63\,GeV. Projected limits for $WW$ (dominant around 
72\,GeV and in the high mass region) are expected to be very similar (see text for details). The CTA projection 
(blue dashed line) assumes annihilation into a $WW$ final state
and an Einasto DM halo profile.
}
\label{fig:IDprosp}
\end{figure}
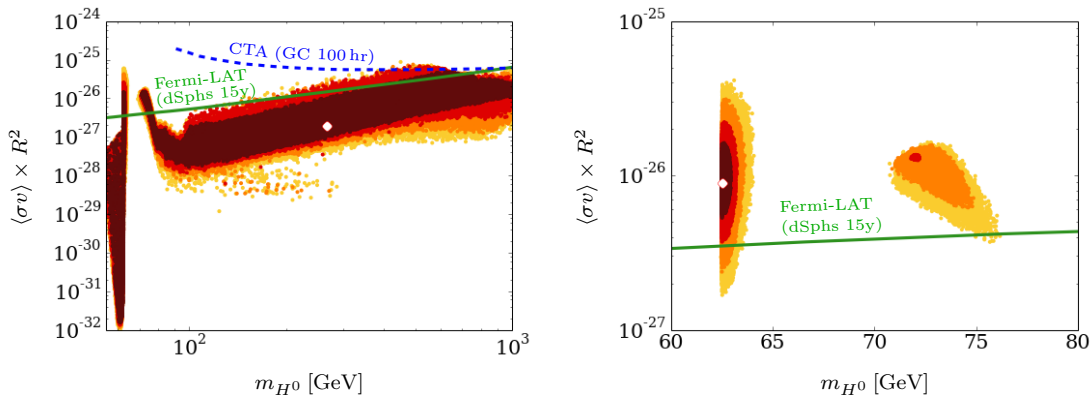

On the side of indirect detection, Fermi-LAT searches for gamma-rays in dwarf spheroidal galaxies will ultimately probe a possible
explanation of the GCE within the IDM\@. To illustrate this point, in figure~\ref{fig:IDprosp} we project our scan points
onto the $m_{\Hn}$-$\sv R^2$ plane and superimpose them with the projected
sensitivity of Fermi-LAT, assuming 15 years of data acquisition~\cite{Charles:2016pgz}.
One subtlety concerns the fact that these projections are derived assuming annihilation into a pure $b\bar b$ final state. 
In the two IDM GCE best-fit regions that we have found
the dominant DM annihilation channels are either $b\bar b$ or $WW^*$, while for the high mass region 
the $WW$ and $ZZ$ channels dominate. Since, however, the current bounds on $\sv$ from dwarf spheroidal galaxies are similar for the $b\bar b$ 
and $WW$ channels (the differences in the derived limits being of the order of 25\%), these projections can, indeed, provide 
an adequate estimate of the indirect detection prospects of the IDM\@. 

The result indicates that the entire 1--2$\sigma$ region that is compatible with a DM interpretation 
of the GCE can be probed with 15 years of LAT data acquisition.

For the high mass region CTA~\cite{Consortium:2010bc} is expected to provide
better sensitivity. In the left panel of figure~\ref{fig:IDprosp} we show the projected
limits for 100hr of CTA observation of the Galactic Center, 
taken from~\cite{Silverwood:2014yza}, assuming annihilation into $WW$ and 
an Einasto DM density profile. Note that for a more cuspy generalized NFW 
profile (as it is found to be compatible with the GCE) those limits are stronger
by up to an order of magnitude~\cite{Silverwood:2014yza}.

%=====================
%    \                                           |
%      \                                         |
%        \                                       |
%                                      \         |
%                                        \       |
%                                          \     |
%=====================

%-   -   -   -   -   -   -   -   -   -   -   -   -   -   -   -   -   -   -
\subsubsection*{Collider searches}
%-   -   -   -   -   -   -   -   -   -   -   -   -   -   -   -   -   -   -
On the side of collider searches, given the very small $\lambda_L$ values that characterise the GCE best-fit regions, the standard LHC mono-X searches for $\Hn$ pairs become inefficient. Of more interest are searches for the heavier ${\cal{Z}}_2$-odd states subsequently decaying into visible products and missing energy. As pointed out in \cite{Belanger:2015kga}, the production of $\An\Hn$ and $\Hp\Hp$ pairs does not depend on $\lambda_L$, as it only involves gauge couplings and could, at least for sufficiently light $\An$ and $\Hp$ masses, give rise to visible signals in the dileptons + MET channel. Larger mass splittings, and in particular once $m_{\An} - m_{\Hn} > m_Z$, are harder to probe due to the kinematic cuts imposed by the corresponding searches in order to elliminate the dominant $Z$-induced background.

Another interesting channel that was suggested in \cite{Belyaev:2016lok} concerns the high-mass region of the IDM and in particular cases where the mass splitting between $\Hn$ and $\Hp$ is small. In this case, CMS searches for disappearing charged tracks at 8 TeV \cite{CMS:2014gxa} already exclude a range of $\Hn$ masses between 490 and 550 GeV for $m_{\Hp} - m_{\Hn} \sim 0.2$ GeV regardless of the value of $\lambda_L$. It would be interesting to follow the evolution of these constraints, since they could provide a rather unique handle on the IDM high-mass regime, especially in cases of small $\lambda_L$ values.

%===================================================================
\section{Conclusion}\label{sec:summary} 
%===================================================================

In this paper we presented a global fit of the IDM taking into account state-of-the-art constraints from collider observables, direct and indirect DM searches as well as theoretical considerations. We performed a detailed exploration of the IDM parameter space and updated upon existing studies that have shown that the low-mass regime of the IDM is by now very efficiently constrained, in particular once direct detection bounds are combined with the LHC Higgs mass measurement and the condition that the IDM reproduces the total DM abundance in the Universe. Going a step further, we relaxed the latter requirement and instead demanded simply that the Universe does not get overclosed assuming a standard thermal history, which allowed us to examine the substantial -- and much less frequently studied -- regions of parameter space that open up.

We then examined whether the IDM can accommodate the excessive Galactic bulge emission that has been reported by numerous groups in the Fermi-LAT data (``Galactic center excess''). We found that this is indeed the case, in two distinct regions of parameter space: the first lies around the so-called ``Higgs funnel'', $m_{\Hn} \sim m_{\hn}/2$, in which $\Hn$ particles annihilate mostly into $b\bar{b}$ pairs through a quasi-on-shell Higgs boson. Interestingly, the strong dependence of the total thermally averaged self-annihilation cross section on the DM velocity in this region of parameter space makes it possible to explain the GCE even if the IDM only accounts for a small fraction of the DM abundance in the Universe, as the cross section computed at velocities relevant for indirect detection can supersede the corresponding ones computed at freeze-out velocities by several orders of magnitude. The second $\Hn$ mass range in which the IDM can explain the GCE lies around 72 GeV, close (but not too close) to the $WW$ threshold. In this case, DM annihilates predominantly into pairs of virtual $W$'s via the quartic $\Hn$-$\Hn$-$W^+$-$W^-$ coupling that appears in the Lagrangian gauge kinetic terms for the inert doublet. In this case the dependence of $\left\langle \sigma v \right\rangle$ on the DM velocity is milder, and the GCE can mostly be explained for $R \equiv \Omega h^2|_\text{IDM} / \Omega h^2|_\text{DM,\,Planck} \sim 1$, which is attainable without conflicting current direct DM searches since the corresponding coupling only enters the WIMP-nucleon scattering cross section at next-to-leading order. To the best of our knowledge, the existence of this region constitutes a novelty of the IDM with respect to simpler ``portal'' models and has not been pointed out before.

Both of these regions of parameter space involve small (${\cal{O}}(10^{-2})$ or less) values of the coupling between $\Hn$ pairs and the standard model Higgs boson and masses $m_{\Hn} \gtrsim m_{\hn}/2$. As a consequence, collider probes of the IDM GCE explanation do not appear to be particularly promising. On the other hand, our findings show that Fermi-LAT constraints from searches for DM in dwarf spheroidal galaxies should confirm or exclude this scenario with 15 years of data acquisition. At the same time, the next generation of direct detection experiments will also provide complementary information in this direction, since the LZ and especially the Darwin experiments will reach a level of sensitivity that will enable them to probe the electroweak radiative corrections to the WIMP-nucleon scattering cross section, which are fixed by the electroweak coupling strength and do not depend, hence, on the additional parameters of the model. Besides, both direct and indirect detection experiments are expected to probe a substantial fraction of the high-mass regime of the IDM, at least for scenarios that saturate the Planck bound on the DM abundance in the Universe.

%===================================================================
\section*{Acknowledgements}
%===================================================================

We would like to thank Alessandro Cuoco and Michael Kr{\"a}mer 
for very helpful discussions. 
B.E. and J.H. acknowledge support by the German Research Foundation DFG through the research unit ``New physics at the LHC''. 

\addcontentsline{toc}{section}{References}
\bibliographystyle{utphys.bst}
\bibliography{GCE_ref}

\end{document}